\documentclass[fleqn,usenatbib]{mnras}

\usepackage{newtxtext,newtxmath}

\usepackage[T1]{fontenc}
\usepackage{ae,aecompl}

\usepackage{graphicx}	
\usepackage{amsmath}	
\usepackage{hyperref}
\usepackage{xspace}
\usepackage{multirow}
\usepackage{verbatim}
\usepackage{soul}

\newcommand{\kms}{\xspace\ensuremath{\rm km\,s^{-1}}}	
\newcommand{\ergcms}{~\ensuremath{\rm erg\,s^{-1}\,cm^{-2}\,\textup{\AA}^{-1}}}
\newcommand{\fluxline}{~\ensuremath{\rm erg\,s^{-1}\,cm^{-2}}}
\newcommand{\ergcmcmcms}{\xspace erg\,cm$^{3}$\,s$^{-1}$}
\newcommand{\zsun}{\xspace\ensuremath{{\rm Z_\odot}}\xspace}
\newcommand{\msun}{\xspace\ensuremath{{\rm M_\odot}}\xspace}

\newcommand{\oiii}{[\ion{O}{iii}]\xspace}
\newcommand{\halpha}{H$\alpha$\xspace}
\newcommand{\hbeta}{H$\beta$\xspace}

\newcommand{\nii}{[\ion{N}{ii}]\xspace}
\newcommand{\sii}{[\ion{S}{ii}]\xspace}
\newcommand{\meanZ}{\ensuremath{\langle {\rm Z} \rangle }\xspace}
\newcommand{\meanAge}{\ensuremath{\langle log \ t_{*}\rangle }\xspace}
\newcommand{\Eout}{\ensuremath{\dot{E}_{out}}\xspace}
\newcommand{\Mout}{\ensuremath{\dot{M}_{out}}\xspace}
\newcommand{\Lbol}{\ensuremath{L_{bol}}\xspace}
\newcommand{\Lhalpha}{\ensuremath{L(H\alpha)}\xspace}

\newcommand{\refappendix}[1]{\hyperref[#1]{Appendix~\ref*{#1}}}

\title[Mapping the Inner kpc of NGC 2992]{
Exploring the AGN-Merger Connection in Arp 245 I: Nuclear Star Formation and Gas Outflow in NGC~2992}

\author[M. Guolo-Pereira et al.]{Muryel Guolo-Pereira$^{1}$\thanks{E-mail:muryel@astro.ufsc.br},
    Daniel Ruschel-Dutra$^{1}$,
   Thaisa Storchi-Bergmann$^{2}$,
    \newauthor
     Allan Schnorr-M\"uller$^{2}$,
   Roberto Cid Fernandes$^{1}$,
    Guilherme Couto$^{3}$,
    \newauthor
    Natacha Dametto$^{3}$,
    Jose A Hernandez-Jimenez$^{4}$
\\
$^{1}$Departamento de F\'isica - CFM - Universidade Federal de Santa Catarina, 476, 88040-900 Florian\'opolis, SC, Brazil\\
$^{2}$Departamento de Astronomia, Universidade Federal do Rio Grande do Sul. Av. Bento Goncalves 9500, 91501-970 Porto Alegre, RS, Brazil\\
$^{3}$Centro de Astronom\'ia (CITEVA), Universidad de Antofagasta, Avenida Angamos 601, Antofagasta, Chile \\
$^{4}$ Departamento de Ciencias F\'isicas, Universidad Andr\'es Bello, Fern\'andez Concha 700, Las Condes, Santiago, Chile
}

\date{Accepted 2021 January 25. Received 2021 January 25; in original form 2019 November 28.}

\pubyear{2021}

\begin{document}
\label{firstpage}
\pagerange{\pageref{firstpage}--\pageref{lastpage}}
\maketitle

\begin{abstract}

Galaxy mergers are central to our understanding of galaxy formation, especially within the context of hierarchical models.
Besides having a large impact on the star formation history, mergers are also able to influence gas motions at the centre of galaxies and trigger an Active Galactic Nucleus (AGN).
In this paper, we present a case study of the Seyfert galaxy NGC~2992, which together with NGC~2993 forms the early-stage merger system Arp 245. Using Gemini Multi-Object Spectrograph (GMOS) integral field unit (IFU) data from the inner  1.1 kpc of the galaxy we were able to spatially resolve the stellar populations, the ionisation mechanism and kinematics of ionised gas.
From full spectral synthesis, we found that the stellar population is primarily composed by old metal-rich stars (t $\geq$ 1.4 Gyr, $Z \geq 2.0$\zsun), with a contribution of at most 30 per cent of the light from a young and metal-poor population (t $\leq$ 100 Myr, $Z \leq 1.0$\zsun).
We detect \halpha and \hbeta emission from the Broad Line Region (BLR) with a Full Width at Half Maximum (FWHM) of $\sim$ 2000\kms.
The Narrow Line Region (NLR) kinematics presents two main components: one from gas orbiting the galaxy disk and a blueshifted (velocity $\approx$ -200\kms) outflow, possibly correlated with the radio emission, with mass outflow rate of $\sim$ 2 M$_{\odot}$ yr$^{-1}$ and a kinematic power of $\sim$ 2 $\times 10^{40}$ erg s$^{-1}$ (\Eout/\Lbol $\approx$ 0.2 per cent).
We also show even though the main ionisation mechanism is the AGN radiation, ionisation by young stars and shocks may also contribute to the emission line ratios presented in the innermost region of the galaxy. 
\end{abstract}




\section{Introduction}
\label{sec:intro}

The vast majority of galaxies with a spheroid host a supermassive black hole (SMBH) in their centres, with a mass range of $\sim10^{6}-10^{10}\,\,{\rm M_\odot}$.
In some galaxies, these objects are active and emit intense radiation due to the accretion of matter onto the SMBH through an inner region called accretion disk \citep{1984RvMP...56..255B, 1998ApJ...501...82P, 2019ApJ...875L...1E}.
Active galactic nuclei (AGN) in Seyfert galaxies are classified as type 1 and 2, depending on the presence of broad emission lines of the Balmer series, with intermediate types between these two also possible; these broad lines are produced in a region called Broad Line Region (BLR).

The importance of AGN in the evolution of the host galaxy is not completely understood, but there is evidence pointing to a co-evolution scenario between the two.
The $M_{\rm BH}-\sigma_{*}$ relation \citep{2000ApJ...539L...9F, 2013ARA&A..51..511K} -- which relates the mass of the SMBH with the stellar velocity dispersion in the galactic bulge -- and the simultaneous peak between star formation rate (SFR) and  SMBH accretion rate in cosmological studies at $z \approx 2$ \citep{2008ApJ...679..118S, 2014ARA&A..52..415M} are some of the indications towards a causal connection between host galaxy evolution and of its SMBH.

The energy released by the AGN in the form of radiation, winds, or radio plasma jets is known to impact on
the interstellar medium of the host galaxy. All these processes are collectively known as feedback. The role of such
AGN feedback in the evolution of galaxies, however, is still a subject of ongoing debate. AGN feedback may be a key factor in quenching star formation, mainly in the late stages of the host galaxy's evolution \citep{2005Natur.433..604D, 2006MNRAS.365...11C}.
In fact, large scale hydrodynamic simulations of galaxy formation are not able to reproduce observed properties of massive galaxies if AGN feedback models are not included \citep{2005Natur.435..629S, 2017ARA&A..55..343B} in which these processes are responsible for suppressing the growth of the most massive galaxies by heating and expelling the gas that would form more stars.
Powerful AGN driven winds are a possible mechanism for preventing further growth of the host galaxy.
These winds can have outflow velocities as high as 1000\kms \citep{Rupke_2011, 2012ApJ...746...86G} and as low as well $\approx 100-200$\kms, with mass outflow rates in nearby galaxies averaging at a few solar masses per year \citep[e.g.][]{Riffel_2011,Crenshaw_2015,Revalski_2018}.

Theoretical studies suggest large scale events like major mergers are the dominant processes leading to SMBH
growth at high masses \citep{2014A&A...569A..37M}. At high redshifts (z > 2), major mergers have also been proposed as
fuelling mechanisms of the fastest-growing SMBHs \citep{Treister_2012}.
A major merger can destabilise large quantities of gas, driving massive inflows towards the nuclear region of galaxies and triggering bursts of star formation and nuclear activity \citep{Hopkins_10,2018MNRAS.479.3952B}.
Many studies have found that the most luminous AGN are preferentially hosted by galaxy mergers \citep{Schawinski_2010,glikman2015major,2016ApJ...822L..32F}.
At lower luminosities, several studies have found a higher incidence of galaxies with signatures of interactions in AGN hosts as compared to control samples \citep{2010ApJ...716L.125K} and particularly in close galaxy pairs \citep{2011MNRAS.418.2043E,2014MNRAS.441.1297S} suggesting kinematic pairs are conducive environments for black hole growth.

One possible consequence of gas inflow is the circumnuclear star formation \citep{Hopkins_08,Hopkins_10}.
Such merger-driven inflows can both increase the star formation rate  \citep[][]{Ellison_2013, Pan_2019} and modify the metallicities gradients of the galaxies \citep{Barrera-Ballesteros_2015}.
However, classical emission-line SFR probes \citep[e.g.][]{Kennicutt_98} and metallicity calibrations \citep[e.g. ][]{Alloin_79,Pettini_04} cannot be applied to active galaxies, due to the contamination of the AGN ionisation to the emission lines.
In this way, the use of stellar populations synthesis, by performing full spectral fitting, can reveal clues on the star formation history (SFH) and chemical evolution of AGN host galaxies and its content.


In this context, we present high spatial and spectral resolution optical IFU observations of the inner 1.1 kpc of NGC~2992. NGC~2992 is a nearby interacting Seyfert galaxy seen almost edge-on \citep[$i\sim70^\circ$,][]{Marquez_1998}  
at a distance, as measured using the Tully-Fisher relation, of $38\,\,{\rm Mpc}$ \citep{Theureau_2007}, which translates into a projected angular scale of $\sim$ 150~pc per arcsec. Alongside with the starburst galaxy NGC~2993 and the tidal dwarf galaxy A245N it forms the system Arp 245. Using hydrodynamic simulations, \cite{Duc_2000} reported the system is currently at an early stage of the interaction, about $\approx 100\,\,{\rm Myr}$ after pericentre passage.
Nevertheless, tidal tails have already developed, which can be seen by optical images (\autoref{fig:Data}), in CO and HI lines \citep{Duc_2000} and infrared (IR) images \citep{2015MNRAS.449.1309G}.

The galaxy nuclear activity has been the subject of several studies, partly due to its variability as seen both in X-rays and in the optical \citep{2008AJ....135.2048T}, even leading to changes in spectral classification.
Early spectra published by \cite{1980ApJ...240...32S}, \cite{1980A&A....87..245V}, and \cite{1980MNRAS.193..563W} all show the presence of a weak broad H$\alpha$ component, originated in the BLR, but no detectable corresponding broad H$\beta$ component, leading to its original classification as a Seyfert 1.9.
Observations from 1994 \citep{allen1999}, however, display no broad \halpha component, thus having the classification of Seyfert 2. \cite{2000A&A...355..485G} reported a broad \halpha emission again, just to disappear once again when observed in 2006 by \cite{Trippe_2008}.
More recently \cite{Schnorr_2016}, after subtracting the narrow component, were able to detect the broad H$\beta$ component for the first time, then classifying the galaxy as a Seyfert 1.8.

In the radio, 6 cm observations show a double lobe 8-shaped structure of about 8 arcsec ($\sim$ 1 kpc) extension to the northwest and Southeast of the galaxy, along position angle (PA) of $-26^\circ$ \citep{Ulvestad_1984}.
From IR observations, \cite{Chapman_2000} suggest the best interpretation is that this structure is related to expanding plasma bubbles, possibly carried by internal jets from the AGN.
More recently, using radio polarimetry, \citet[][]{Irwin_2017} found another double-lobed radio morphology within its spiral disc, which was revealed in linearly
polarized emission but not in total intensity emission. This second structure by \citet[][]{Irwin_2017} is much more extended than the one found by \citet{Ulvestad_1984} reaching several kpcs from the nucleus, being interpreted by the authors as a relic of an earlier episode of AGN activity.

NGC~2992 gas  kinematics is complex as found by several long slit spectroscopy studies \citep[e.g. ][]{Marquez_1998, Veilleux_2001} as well as by several Integral Field Unit observations \citep[][]{Garcia_lorenzo_01, Friedrich_2010,M_sanchez_2011}.
These observations show, at most positions, the presence of a double component line profile.
While one component follows the galaxy rotation curve, the other is interpreted as outflowing gas by the authors. Using The Multi Unit Spectroscopic Explorer \citep[MUSE][]{Bacon_2010} at the Very Larga Telescope (VLT) \citet[][]{Mingozzi_2019} clearly show the presence of a kpc-scale, bipolar outflow with a large opening angle.

This paper is the first in a series which will explore the possible connection between merger and AGN in Arp\,245.
In this first paper, we analyse both the stellar populations and the properties of the ionised gas of NGC~2992, in order to better understand the physical conditions of the nuclear region and the mutual role of nuclear activity and the interaction may play.
In a second paper (Guolo-Pereira et al., in prep., Paper~II) we will observationally explore the effects of the interaction in NGC~2993.
While in the third one (L\"osch et al., in prep, Paper~III) we will present a modern version of the hydrodynamics simulations by \cite{Duc_2000} focusing on the possible triggering of the AGN during the merger.

The present paper is organised as follows: in \autoref{sec:observations} we describe the observations and data
reduction; \autoref{sec:stellar} we present the stellar populations synthesis and its spatial distribution; the ionised
gas kinematics is described in \autoref{sec:kinematics}; ionisation mechanisms are discussed in \autoref{sec:Ion_Mec}; \autoref{sec:discussion} presents our discussion of the main results and their significance; and finally
\autoref{sec:conclusion} contains our conclusions.

\section{Observations and data reduction}
\label{sec:observations}

\begin{figure*}
    \centering
	\includegraphics[width=0.7\textwidth]{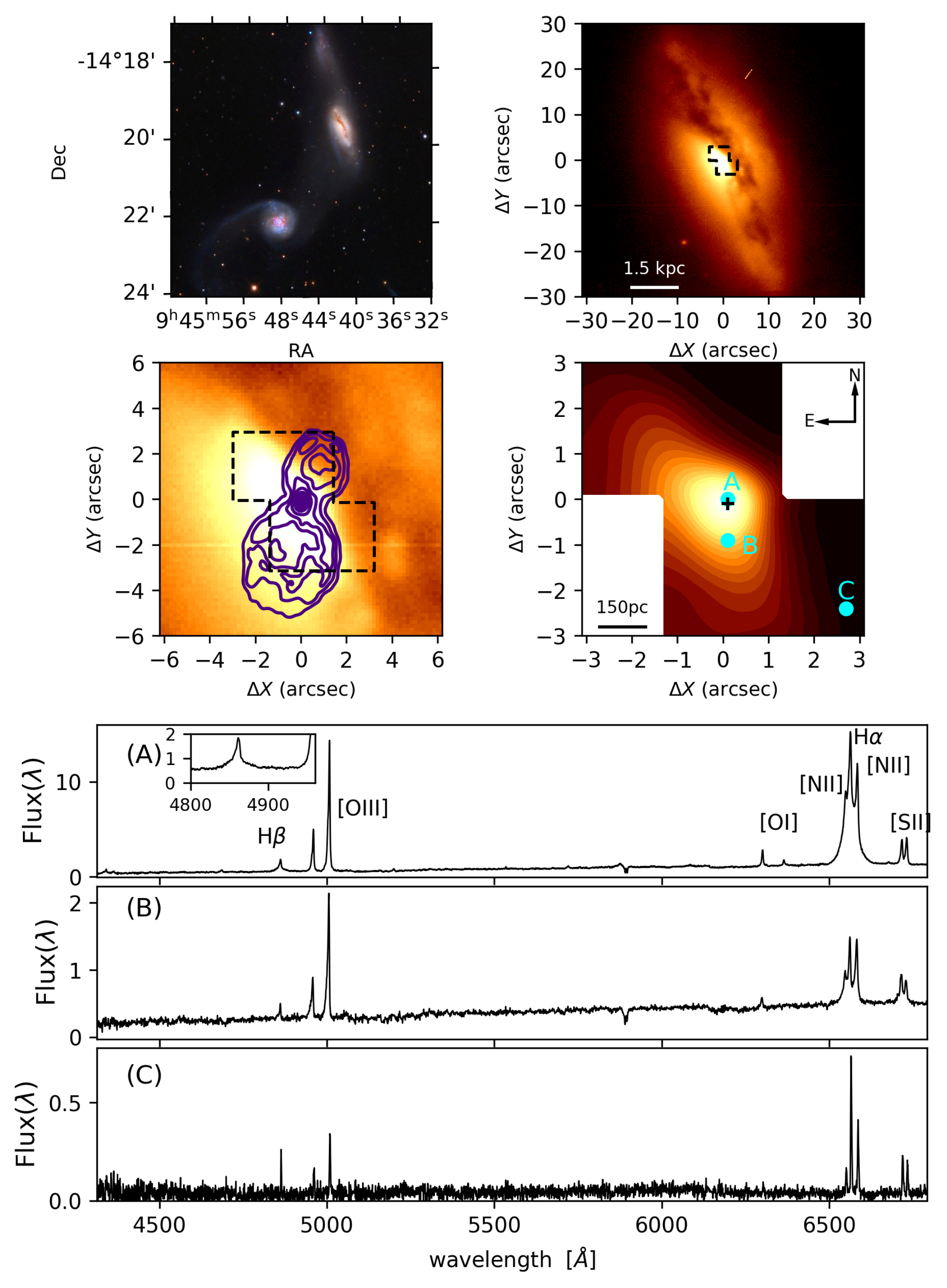}
    \caption{Top Left: LRGB image composition of Arp 245 from           \citet{Block_2011}. Top Right: GMOS r' band               acquisition image of NGC 2992. Middle Left: Zoomed         acquisition image, and the
        \citet{Ulvestad_1984}  radio emission in purple (contours : 3, 6, 9, 15, 30, 50, 70, 90 per cent
        of the peak of 7.2 mJy). The dashed rectangles show the two IFU FoV's. Middle Right: IFU continuum mean flux in the FoV. The black cross represents the continuum nuclear position in all figures of the paper. All maps of the paper were smoothed using a Gaussian kernel with FHWM with half of the spatial resolution, see \autoref{sec:observations}. Bottom, from top to bottom: Spectra corresponding to the regions marked as A, B and C, with spectral flux density in units of $10^{-15}$\ergcms. 
        North is up and East to the left in this and all figures of the paper.
    }
    \label{fig:Data}
\end{figure*}

This work is mostly based on data obtained with the Integral Field Unit (IFU) of the Gemini Multi Object Spectrograph (GMOS) at the Gemini South telescope on the night of February 15, 2018 (Gemini project GS-2018A-Q-208, P.I. Ruschel-Dutra).
The observations consisted of two adjacent IFU fields (covering 5\arcsec $\times$ 3\farcs5 each).
The fields are slightly offset, both in Right Ascension (RA) and Declination (Dec), in order to increase the field of view (FoV) and cover a larger portion of the galaxy plane.
For each IFU field, there are six exposures, with an integration time of 560 seconds each.
Each science observation is accompanied by a  5 $\times$ 1\farcs75 sky observation for atmospheric lines removal.

The spatial resolution is limited by seeing, which was at $0\farcs8$ during the observations, corresponding to 120~pc at the galaxy.
The observed spectra covered a range from 4075 to 7285 $\textup{\AA}$ with an instrumental dispersion $\sigma_{\rm instr} \approx 40\,\,\kms$, obtained from the Copper-Argon arc lamp lines.

The reduction procedures were performed using the Gmos Ifu REDuction Suite (GIREDS)\footnote{
    GIREDS is a set of python scripts to automate the reduction of GMOS IFU data, available at
    \href{https://github.com/danielrd6/gireds}{GitHub}.
    }.
The process comprised the usual steps: bias subtraction, flatfielding, trimming, wavelength calibration, telluric emission subtraction, relative flux calibration, the building of the data cubes, and finally the registering and combination of the 12 individual exposures.
The final cube was built with a spatial sampling of $0\farcs1 \times 0\farcs1$, resulting in a FoV with $\approx 2800$ spaxels. After cropping the borders and superimposing the FoVs, it results in a total useful combined FoV, with total angular coverage of 29~arcsec{$^2$} around the nucleus. The combined FoV is centred at RA = 9h45m42s and Dec = -14$^{o}$9\arcmin3\arcsec, while the continuum peak is shifted at $\Delta$RA = 0.1\arcsec and $\Delta$Dec = -0.1\arcsec.

Many Gemini data cubes show certain ``instrumental fingerprints'', in the form of vertical stripes in the reconstructed data cube.
Such features also have specific spectral signatures.
In order to identify and remove these instrumental features, we use a technique based on Principal Component Analysis \citep[PCA, ][]{2014MNRAS.438.2597M}.
The fingerprint removal is fundamental for the stellar population synthesis because its effect is more apparent for lower instrumental counts.
The data cube spectra were corrected for Galactic extinction using the \citet{1989ApJ...345..245C} extinction curve (CCM) for an $A_V=0.166$ \citep{Schlafly_11} and by Doppler effect using the radial velocity of 2345\kms (see \autoref{sec:kinematics}).

We show in the upper panel of the \autoref{fig:Data} an LRGB image of the Arp 245 system (Left) and the acquisition image taken with a r' filter (Right). In the middle panel we show a zoomed version of the acquisition image superimposed with the the \cite{Ulvestad_1984} figure of 8-shaped radio emission of the galaxy (Left) and the mean continuum flux in the IFU's
FoV (Right), while in the bottom panels we show three sample spectra from representative regions marked, respectively from top to bottom, A, B and C.
The prominent dust lane seen in the acquisition image, and in the stellar continuum image from the IFU, has a
reported $A_V$ greater than 3 \citep{1987A&A...178...51C, Schnorr_2016}. 
Spectra from this region of high extinction have no measurable stellar component above the noise level, as exemplified
by the spectrum extracted from C.
The nuclear
spectrum shows a broad  H${\alpha}$ component, and a faint broad H$\beta$ is also present, in contrast with its original
Seyfert 1.9 classification but in agreement with its the current state, Seyfert 1.8, proposed by
\cite{Schnorr_2016}.

\section{Stellar Populations}
\label{sec:stellar}

In order to obtain the spatially resolved SFH, and to fit the gas emission features, free from the stellar population contamination, we performed stellar population synthesis on the data cube by employing the \textsc{starlight} full spectra fitting code \citep{Cid_2005}.


Briefly, \textsc{starlight} fits an observed stellar spectrum $O_{\lambda}$ with a model $M_\lambda$ built from a linear combination of $N$ spectral components.
Dust attenuation is treated as if due to a foreground screen and parametrised by the $V$-band extinction $A_V$.
Stellar kinematics is modelled with a Gaussian line-of-sight velocity distribution centred at velocity $v_*$, and with a dispersion $\sigma_*$.
Besides the model spectrum $M_\lambda$, the code outputs the corresponding values of $A_V$, $v_\star$, $\sigma_\star$, and a population vector $\Vec{x}$ containing the percentage contribution of each base population to the flux at a chosen normalisation wavelength $\lambda_0$, which in our fits was chosen as $\lambda_0$ = 5500 $\textup{\AA}$. 
The code minimises the $\chi^2$ of the fit, though a more convenient and intuitive figure of merit to assess the quality of the fit is through \textit{adev} parameter, defined as the mean value of $\left |O_{\lambda}  - M_{\lambda} \right |/O_{\lambda}$ over the fitted wavelengths.

We note that, while there are many different spectral fitting codes and techniques available, cross-comparison of these various techniques have shown to yield rather consistent results \citep[e.g. ][]{Koleva_2008,Mateu_2012,Maksym_2014,Mentz_2016, Cid_Fernandes_2018}.
Therefore, rather than trying different fitting codes, we chose to maintain {\sc starlight} and investigate the differences introduced by the choice of models and stellar libraries, which are known to produce variations in the recovered properties \citep[e.g.][]{Maraston_2011,Chen_2010,Wilkinson_2017,Baldwin_2018,Dametto_2019}.

\subsection{Evolutionary Models, Stellar Libraries and Featureless Continuum}
\label{sec:ssps_and_fc}

This section discusses the stellar population models and stellar libraries used in this work, and how we can account for the contribution of a featureless continuum due to the AGN in NGC~2992.

\defcitealias{BC_2003}{BC03}
\defcitealias{Maraston_2011}{M11}

The key ingredients in the stellar population synthesis method are the spectral base elements, i.e. those spectra available to the fitting code to combine in order to recover the galaxy properties.
The base elements are, usually, Simple Stellar Population (SSP), i.e. a collection of stars with same age and metallicity, taken from an evolutionary model which are constructed using a stellar spectra library and an Initial Mass Function (IMF).

In this work we have used SSPs  from two models: \citet[][BC03]{BC_2003}, which uses the STELIB \citep{2003A&A...402..433L} spectral library,
and \citet[][M11]{Maraston_2011}, which employs the MILES \citep{Miles_2006} spectral library.
For both models/libraries we use the same \cite{2003PASP..115..763C} IMF and instantaneous star formation bursts.
We found varying the IMF produces only minor variations in the recovered SFH when compared to the changes introduced by using distinct models. The former is outside the scope of this work, and we refer the reader interested in this particular effect to \citet{Chen_2010}, \citet{Ge_2019} and references therein.
From the SSPs provided by \citetalias{BC_2003} we choose a set of 45 representative elements divided into three metallicities ($0.2, 1.0, 2.5$\zsun) and an age range covering from t(Gyr) $= 0.001$ to t(Gyr) $= 13$.
From those available by  \citetalias{Maraston_2011} with the MILES library we chose a set of 40 SSPs with age and metallicity coverage as similar as possible to the \citetalias{BC_2003} ones: three metallicities ($0.5, 1.0, 2.0$\zsun) and ages ranging from  t(Gyr) $= 0.007$ to t(Gyr) $ = 13$.
The complete list of SSPs used from each model is shown in \autoref{table:SSPs}.
It is important to mention that the MILES Library does not provide metal-poor stars ($Z = 0.5$\zsun) younger than 55 Myr (0.055 Gyr), and neither metal-rich stars ($Z = 2.0$\zsun) younger than 100 Myr (0.1 Gyr), while \citetalias{BC_2003}/STELIB provide the same age range for the three metallicities (see \autoref{table:SSPs}).

We define a reduced population vector with three age ranges: $t < $ 100 \ Myr; \ 100\ Myr  \ $\leq t \leq$ 1.4   Gyr;
and $ \ t  >  1.4 $~Gyr,  to obtain a more robust description of the SFH than that obtained with the full population vector \citep[see, e.g.][]{Cid_2005}, as well as to facilitate the  visualisation of the spatially resolved fit results. The population vector components are denoted by $x_Y$, $x_I$ and $x_O$, where $`Y'$, $`I'$ and $`O'$ stands for young,
intermediate and old respectively. The division in a reduced population vector has been used in several works in stellar population synthesis \citep[e.g.][]{Cid_2005,Riffel_2010,Cid_2013,Gonzalez_Delgado_2015,Mallmann_2018}, while the range values may be slightly distinct from one another, the ones adopted here are the most common. Also, for the young population the cutting value of 100 Myr is the estimated age of the pericentre passage between NGC2992 and its companion \citep[][]{Duc_2000}.

\begin{table*}
\caption{SSPs used in the stellar population synthesis.}
\centering
\begin{tabular}{|c|c|c|c|c|c|c|} 
\hline
Model         & \multicolumn{3}{c|}{BC03}                                                                                                                                                                                                                                                                                                                                                                                                                                                                                                                                                                                                                                                                                                                                                                                                                                                                                                                                                                                                                                                                                                                                                                                                                              & \multicolumn{3}{c|}{M11}                                                                                                                                                                                                                                                                                                                                                                                                                                                                                                                                                                                                                                                                                                                                                                                                                                                                                                                                                                                                                                                                                                                                                                                                                                                                                                                                                                                                                                                                                                                                                                                                                                                                                                                                                                                                                                                                                                                                                                                                                                                                                                                                                                                                                                                                                                                                                                                                                                                                                                                                                                                                                                                                                                                                                                                                                                                                                                                                                                                                                                                                                                                                                                                                                                                                                                                                                                                                                                                                                                                                                                                                                                                                                                                                                                                                                                                                                                                                                                                                                                                                                                                                                                                                                                                                                                                                                                                                                                                                                                                                                                                                                                                                                                                                                                                                                                                                                                                                                                                                                                                                                                                                                                                                                                                                                                                                                                                                                                                                                                                                                                                                                                                                                                                                                                                                                                                                                                                                                                                                                                                                                                                                                                                                                                                                                                                                                                                                                                                                                                                                                                                                                                                                                                                                                                                                                                                                                                                                                                                                                                                                                                                                                                                                                                                                                                                                                                                                                                                                                                                                                                                                                                                                                                        \\ 
\hline
Spectral Library  & \multicolumn{3}{c|}{STELIB}                                                                                                                                                                                                                                                                                                                                                                                                                                                                                                                                                                                                                                                                                                                                                                                                                                                                                                                                                                                                                                                                                                                                                                                                                            & \multicolumn{3}{c|}{MILES}                                                                                                                                                                                                                                                                                                                                                                                                                                                                                                                                                                                                                                                                                                                                                                                                                                                                                                                                                                                                                                                                                                                                                                                                                                                                                                                                                                                                                                                                                                                                                                                                                                                                                                                                                                                                                                                                                                                                                                                                                                                                                                                                                                                                                                                                                                                                                                                                                                                                                                                                                                                                                                                                                                                                                                                                                                                                                                                                                                                                                                                                                                                                                                                                                                                                                                                                                                                                                                                                                                                                                                                                                                                                                                                                                                                                                                                                                                                                                                                                                                                                                                                                                                                                                                                                                                                                                                                                                                                                                                                                                                                                                                                                                                                                                                                                                                                                                                                                                                                                                                                                                                                                                                                                                                                                                                                                                                                                                                                                                                                                                                                                                                                                                                                                                                                                                                                                                                                                                                                                                                                                                                                                                                                                                                                                                                                                                                                                                                                                                                                                                                                                                                                                                                                                                                                                                                                                                                                                                                                                                                                                                                                                                                                                                                                                                                                                                                                                                                                                                                                                                                                                                                                                                                      \\ 
\hline
Metallicity ($Z_{\odot}$) & 0.2 & 1.0 & 2.5                                                                                                                                                                                                                                                                                                                                                                                                                                                                                                                                                                                                                                                                                                                                                                                                                                                                                                                                                                                                                                                                                                                                                                                                                                        & 0.5                                                                                                                                                                                                                                                                                                                                                                                                                                                                                                                                                                                                                                                                                                                                                                                                                                                                                                                                                                                                                                                                                                                                                                                             & 1.0                                                                                                                                                                                                                                                                                                                                                                                                                                                                                                                                                                                                                                                                                                                                                                                                                                                                                                                                                                                                                                                                                                                                                                                                                                                                                                                                                                                                                                                                                                                                                                                                                                                                                                                                                                                                                                                                                                                                                                                                                                                                                                                                                                                                                                                                                                                                                                                                                                                                                                                                                                                                                                                                                                                                                                                                                                                                                                                                                                                                                                                                                                                                                                                                                                                                                                                                                                                                                                                                                                                                                                                                                                                                                                                                                                                                                                                                                                                                                                                                                                                                                                                                                                                                                                                                                                                                                                                                                                                                                                                                                                                                                                                                                                                                                                                                                                                                                                                                                                                                                                                                                                                                                                                                                                                                                                                                & 2.0                                                                                                                                                                                                                                                                                                                                                                                                                                                                                                                                                                                                                                                                                                                                                                                                                                                                                                                                                                                                                                                                                                                                                                                      \\ 
\hline
Ages (Gyr)        & \multicolumn{3}{c|}{\begin{tabular}[c]{@{}c@{}}\textcolor[rgb]{0.2,0.2,0.2}{0.001, 0.003, 0.005, }\\\textcolor[rgb]{0.2,0.2,0.2}{\textcolor[rgb]{0.2,0.2,0.2}{\textcolor[rgb]{0.2,0.2,0.2}{0.01, 0.025, ~0.04, }}}\\\textcolor[rgb]{0.2,0.2,0.2}{\textcolor[rgb]{0.2,0.2,0.2}{\textcolor[rgb]{0.2,0.2,0.2}{\textcolor[rgb]{0.2,0.2,0.2}{\textcolor[rgb]{0.2,0.2,0.2}{\textcolor[rgb]{0.2,0.2,0.2}{0.1, 0.3, 0.6, 0.9, }}}}}}\\\textcolor[rgb]{0.2,0.2,0.2}{\textcolor[rgb]{0.2,0.2,0.2}{\textcolor[rgb]{0.2,0.2,0.2}{\textcolor[rgb]{0.2,0.2,0.2}{\textcolor[rgb]{0.2,0.2,0.2}{\textcolor[rgb]{0.2,0.2,0.2}{\textcolor[rgb]{0.2,0.2,0.2}{\textcolor[rgb]{0.2,0.2,0.2}{\textcolor[rgb]{0.2,0.2,0.2}{\textcolor[rgb]{0.2,0.2,0.2}{1.5, 3.0, 5.0, }}}}}}}}}}\\\textcolor[rgb]{0.2,0.2,0.2}{\textcolor[rgb]{0.2,0.2,0.2}{\textcolor[rgb]{0.2,0.2,0.2}{\textcolor[rgb]{0.2,0.2,0.2}{\textcolor[rgb]{0.2,0.2,0.2}{\textcolor[rgb]{0.2,0.2,0.2}{\textcolor[rgb]{0.2,0.2,0.2}{\textcolor[rgb]{0.2,0.2,0.2}{\textcolor[rgb]{0.2,0.2,0.2}{\textcolor[rgb]{0.2,0.2,0.2}{\textcolor[rgb]{0.2,0.2,0.2}{\textcolor[rgb]{0.2,0.2,0.2}{\textcolor[rgb]{0.2,0.2,0.2}{\textcolor[rgb]{0.2,0.2,0.2}{\textcolor[rgb]{0.2,0.2,0.2}{11.0, 13.0} }}}}}}}}}}}}}}\end{tabular}} & \begin{tabular}[c]{@{}c@{}}\textcolor[rgb]{0.2,0.2,0.2}{0.055, 0.065, }\\\textcolor[rgb]{0.2,0.2,0.2}{\textcolor[rgb]{0.2,0.2,0.2}{\textcolor[rgb]{0.2,0.2,0.2}{0.075,~0.085,}}}\\\textcolor[rgb]{0.2,0.2,0.2}{\textcolor[rgb]{0.2,0.2,0.2}{\textcolor[rgb]{0.2,0.2,0.2}{\textcolor[rgb]{0.2,0.2,0.2}{\textcolor[rgb]{0.2,0.2,0.2}{\textcolor[rgb]{0.2,0.2,0.2}{~0.1, 0.15,}}}}}}\\\textcolor[rgb]{0.2,0.2,0.2}{\textcolor[rgb]{0.2,0.2,0.2}{\textcolor[rgb]{0.2,0.2,0.2}{\textcolor[rgb]{0.2,0.2,0.2}{\textcolor[rgb]{0.2,0.2,0.2}{\textcolor[rgb]{0.2,0.2,0.2}{\textcolor[rgb]{0.2,0.2,0.2}{\textcolor[rgb]{0.2,0.2,0.2}{\textcolor[rgb]{0.2,0.2,0.2}{\textcolor[rgb]{0.2,0.2,0.2}{~0.3, 0.6, }}}}}}}}}}\\\textcolor[rgb]{0.2,0.2,0.2}{\textcolor[rgb]{0.2,0.2,0.2}{\textcolor[rgb]{0.2,0.2,0.2}{\textcolor[rgb]{0.2,0.2,0.2}{\textcolor[rgb]{0.2,0.2,0.2}{\textcolor[rgb]{0.2,0.2,0.2}{\textcolor[rgb]{0.2,0.2,0.2}{\textcolor[rgb]{0.2,0.2,0.2}{\textcolor[rgb]{0.2,0.2,0.2}{\textcolor[rgb]{0.2,0.2,0.2}{\textcolor[rgb]{0.2,0.2,0.2}{\textcolor[rgb]{0.2,0.2,0.2}{\textcolor[rgb]{0.2,0.2,0.2}{\textcolor[rgb]{0.2,0.2,0.2}{\textcolor[rgb]{0.2,0.2,0.2}{0.9} }}}}}}}}}}}}}}\end{tabular} & \begin{tabular}[c]{@{}c@{}}\textcolor[rgb]{0.2,0.2,0.2}{0.007, 0008, }\\\textcolor[rgb]{0.2,0.2,0.2}{\textcolor[rgb]{0.2,0.2,0.2}{\textcolor[rgb]{0.2,0.2,0.2}{0.009, 0.01, }}}\\\textcolor[rgb]{0.2,0.2,0.2}{\textcolor[rgb]{0.2,0.2,0.2}{\textcolor[rgb]{0.2,0.2,0.2}{\textcolor[rgb]{0.2,0.2,0.2}{\textcolor[rgb]{0.2,0.2,0.2}{\textcolor[rgb]{0.2,0.2,0.2}{0.015, 0.025,}}}}}}\\\textcolor[rgb]{0.2,0.2,0.2}{\textcolor[rgb]{0.2,0.2,0.2}{\textcolor[rgb]{0.2,0.2,0.2}{\textcolor[rgb]{0.2,0.2,0.2}{\textcolor[rgb]{0.2,0.2,0.2}{\textcolor[rgb]{0.2,0.2,0.2}{\textcolor[rgb]{0.2,0.2,0.2}{\textcolor[rgb]{0.2,0.2,0.2}{\textcolor[rgb]{0.2,0.2,0.2}{\textcolor[rgb]{0.2,0.2,0.2}{0.04, 0.055,}}}}}}}}}}\\\textcolor[rgb]{0.2,0.2,0.2}{\textcolor[rgb]{0.2,0.2,0.2}{\textcolor[rgb]{0.2,0.2,0.2}{\textcolor[rgb]{0.2,0.2,0.2}{\textcolor[rgb]{0.2,0.2,0.2}{\textcolor[rgb]{0.2,0.2,0.2}{\textcolor[rgb]{0.2,0.2,0.2}{\textcolor[rgb]{0.2,0.2,0.2}{\textcolor[rgb]{0.2,0.2,0.2}{\textcolor[rgb]{0.2,0.2,0.2}{\textcolor[rgb]{0.2,0.2,0.2}{\textcolor[rgb]{0.2,0.2,0.2}{\textcolor[rgb]{0.2,0.2,0.2}{\textcolor[rgb]{0.2,0.2,0.2}{\textcolor[rgb]{0.2,0.2,0.2}{0.065, 0.075,}}}}}}}}}}}}}}}\\\textcolor[rgb]{0.2,0.2,0.2}{\textcolor[rgb]{0.2,0.2,0.2}{\textcolor[rgb]{0.2,0.2,0.2}{\textcolor[rgb]{0.2,0.2,0.2}{\textcolor[rgb]{0.2,0.2,0.2}{\textcolor[rgb]{0.2,0.2,0.2}{\textcolor[rgb]{0.2,0.2,0.2}{\textcolor[rgb]{0.2,0.2,0.2}{\textcolor[rgb]{0.2,0.2,0.2}{\textcolor[rgb]{0.2,0.2,0.2}{\textcolor[rgb]{0.2,0.2,0.2}{\textcolor[rgb]{0.2,0.2,0.2}{\textcolor[rgb]{0.2,0.2,0.2}{\textcolor[rgb]{0.2,0.2,0.2}{\textcolor[rgb]{0.2,0.2,0.2}{\textcolor[rgb]{0.2,0.2,0.2}{\textcolor[rgb]{0.2,0.2,0.2}{\textcolor[rgb]{0.2,0.2,0.2}{\textcolor[rgb]{0.2,0.2,0.2}{\textcolor[rgb]{0.2,0.2,0.2}{\textcolor[rgb]{0.2,0.2,0.2}{0.085, 0.1, }}}}}}}}}}}}}}}}}}}}}\\\textcolor[rgb]{0.2,0.2,0.2}{\textcolor[rgb]{0.2,0.2,0.2}{\textcolor[rgb]{0.2,0.2,0.2}{\textcolor[rgb]{0.2,0.2,0.2}{\textcolor[rgb]{0.2,0.2,0.2}{\textcolor[rgb]{0.2,0.2,0.2}{\textcolor[rgb]{0.2,0.2,0.2}{\textcolor[rgb]{0.2,0.2,0.2}{\textcolor[rgb]{0.2,0.2,0.2}{\textcolor[rgb]{0.2,0.2,0.2}{\textcolor[rgb]{0.2,0.2,0.2}{\textcolor[rgb]{0.2,0.2,0.2}{\textcolor[rgb]{0.2,0.2,0.2}{\textcolor[rgb]{0.2,0.2,0.2}{\textcolor[rgb]{0.2,0.2,0.2}{\textcolor[rgb]{0.2,0.2,0.2}{\textcolor[rgb]{0.2,0.2,0.2}{\textcolor[rgb]{0.2,0.2,0.2}{\textcolor[rgb]{0.2,0.2,0.2}{\textcolor[rgb]{0.2,0.2,0.2}{\textcolor[rgb]{0.2,0.2,0.2}{\textcolor[rgb]{0.2,0.2,0.2}{\textcolor[rgb]{0.2,0.2,0.2}{\textcolor[rgb]{0.2,0.2,0.2}{\textcolor[rgb]{0.2,0.2,0.2}{\textcolor[rgb]{0.2,0.2,0.2}{\textcolor[rgb]{0.2,0.2,0.2}{\textcolor[rgb]{0.2,0.2,0.2}{0.15, 0.3, 0.9,}}}}}}}}}}}}}}}}}}}}}}}}}}}}\\\textcolor[rgb]{0.2,0.2,0.2}{\textcolor[rgb]{0.2,0.2,0.2}{\textcolor[rgb]{0.2,0.2,0.2}{\textcolor[rgb]{0.2,0.2,0.2}{\textcolor[rgb]{0.2,0.2,0.2}{\textcolor[rgb]{0.2,0.2,0.2}{\textcolor[rgb]{0.2,0.2,0.2}{\textcolor[rgb]{0.2,0.2,0.2}{\textcolor[rgb]{0.2,0.2,0.2}{\textcolor[rgb]{0.2,0.2,0.2}{\textcolor[rgb]{0.2,0.2,0.2}{\textcolor[rgb]{0.2,0.2,0.2}{\textcolor[rgb]{0.2,0.2,0.2}{\textcolor[rgb]{0.2,0.2,0.2}{\textcolor[rgb]{0.2,0.2,0.2}{\textcolor[rgb]{0.2,0.2,0.2}{\textcolor[rgb]{0.2,0.2,0.2}{\textcolor[rgb]{0.2,0.2,0.2}{\textcolor[rgb]{0.2,0.2,0.2}{\textcolor[rgb]{0.2,0.2,0.2}{\textcolor[rgb]{0.2,0.2,0.2}{\textcolor[rgb]{0.2,0.2,0.2}{\textcolor[rgb]{0.2,0.2,0.2}{\textcolor[rgb]{0.2,0.2,0.2}{\textcolor[rgb]{0.2,0.2,0.2}{\textcolor[rgb]{0.2,0.2,0.2}{\textcolor[rgb]{0.2,0.2,0.2}{\textcolor[rgb]{0.2,0.2,0.2}{\textcolor[rgb]{0.2,0.2,0.2}{\textcolor[rgb]{0.2,0.2,0.2}{\textcolor[rgb]{0.2,0.2,0.2}{\textcolor[rgb]{0.2,0.2,0.2}{\textcolor[rgb]{0.2,0.2,0.2}{\textcolor[rgb]{0.2,0.2,0.2}{\textcolor[rgb]{0.2,0.2,0.2}{\textcolor[rgb]{0.2,0.2,0.2}{1.5, 3.0, 5.0,}}}}}}}}}}}}}}}}}}}}}}}}}}}}}}}}}}}}\\\textcolor[rgb]{0.2,0.2,0.2}{\textcolor[rgb]{0.2,0.2,0.2}{\textcolor[rgb]{0.2,0.2,0.2}{\textcolor[rgb]{0.2,0.2,0.2}{\textcolor[rgb]{0.2,0.2,0.2}{\textcolor[rgb]{0.2,0.2,0.2}{\textcolor[rgb]{0.2,0.2,0.2}{\textcolor[rgb]{0.2,0.2,0.2}{\textcolor[rgb]{0.2,0.2,0.2}{\textcolor[rgb]{0.2,0.2,0.2}{\textcolor[rgb]{0.2,0.2,0.2}{\textcolor[rgb]{0.2,0.2,0.2}{\textcolor[rgb]{0.2,0.2,0.2}{\textcolor[rgb]{0.2,0.2,0.2}{\textcolor[rgb]{0.2,0.2,0.2}{\textcolor[rgb]{0.2,0.2,0.2}{\textcolor[rgb]{0.2,0.2,0.2}{\textcolor[rgb]{0.2,0.2,0.2}{\textcolor[rgb]{0.2,0.2,0.2}{\textcolor[rgb]{0.2,0.2,0.2}{\textcolor[rgb]{0.2,0.2,0.2}{\textcolor[rgb]{0.2,0.2,0.2}{\textcolor[rgb]{0.2,0.2,0.2}{\textcolor[rgb]{0.2,0.2,0.2}{\textcolor[rgb]{0.2,0.2,0.2}{\textcolor[rgb]{0.2,0.2,0.2}{\textcolor[rgb]{0.2,0.2,0.2}{\textcolor[rgb]{0.2,0.2,0.2}{\textcolor[rgb]{0.2,0.2,0.2}{\textcolor[rgb]{0.2,0.2,0.2}{\textcolor[rgb]{0.2,0.2,0.2}{\textcolor[rgb]{0.2,0.2,0.2}{\textcolor[rgb]{0.2,0.2,0.2}{\textcolor[rgb]{0.2,0.2,0.2}{\textcolor[rgb]{0.2,0.2,0.2}{\textcolor[rgb]{0.2,0.2,0.2}{\textcolor[rgb]{0.2,0.2,0.2}{\textcolor[rgb]{0.2,0.2,0.2}{\textcolor[rgb]{0.2,0.2,0.2}{\textcolor[rgb]{0.2,0.2,0.2}{\textcolor[rgb]{0.2,0.2,0.2}{\textcolor[rgb]{0.2,0.2,0.2}{\textcolor[rgb]{0.2,0.2,0.2}{\textcolor[rgb]{0.2,0.2,0.2}{\textcolor[rgb]{0.2,0.2,0.2}{11.0, 13.0} }}}}}}}}}}}}}}}}}}}}}}}}}}}}}}}}}}}}}}}}}}}}\end{tabular} & \begin{tabular}[c]{@{}c@{}}\textcolor[rgb]{0.2,0.2,0.2}{0.1, 0.15,}\\\textcolor[rgb]{0.2,0.2,0.2}{\textcolor[rgb]{0.2,0.2,0.2}{\textcolor[rgb]{0.2,0.2,0.2}{~0.3, 0.6,}}}\\\textcolor[rgb]{0.2,0.2,0.2}{\textcolor[rgb]{0.2,0.2,0.2}{\textcolor[rgb]{0.2,0.2,0.2}{\textcolor[rgb]{0.2,0.2,0.2}{\textcolor[rgb]{0.2,0.2,0.2}{\textcolor[rgb]{0.2,0.2,0.2}{0.9, 1.5, }}}}}}\\\textcolor[rgb]{0.2,0.2,0.2}{\textcolor[rgb]{0.2,0.2,0.2}{\textcolor[rgb]{0.2,0.2,0.2}{\textcolor[rgb]{0.2,0.2,0.2}{\textcolor[rgb]{0.2,0.2,0.2}{\textcolor[rgb]{0.2,0.2,0.2}{\textcolor[rgb]{0.2,0.2,0.2}{\textcolor[rgb]{0.2,0.2,0.2}{\textcolor[rgb]{0.2,0.2,0.2}{\textcolor[rgb]{0.2,0.2,0.2}{3.0, 5.0,}}}}}}}}}}\\\textcolor[rgb]{0.2,0.2,0.2}{\textcolor[rgb]{0.2,0.2,0.2}{\textcolor[rgb]{0.2,0.2,0.2}{\textcolor[rgb]{0.2,0.2,0.2}{\textcolor[rgb]{0.2,0.2,0.2}{\textcolor[rgb]{0.2,0.2,0.2}{\textcolor[rgb]{0.2,0.2,0.2}{\textcolor[rgb]{0.2,0.2,0.2}{\textcolor[rgb]{0.2,0.2,0.2}{\textcolor[rgb]{0.2,0.2,0.2}{\textcolor[rgb]{0.2,0.2,0.2}{\textcolor[rgb]{0.2,0.2,0.2}{\textcolor[rgb]{0.2,0.2,0.2}{\textcolor[rgb]{0.2,0.2,0.2}{\textcolor[rgb]{0.2,0.2,0.2}{11.0, 13.0} }}}}}}}}}}}}}}\end{tabular}  \\
\hline
\label{table:SSPs}
\end{tabular}
\end{table*}
When dealing with stellar populations in Seyfert galaxies, a power law ($F_\lambda \propto\lambda^{- \alpha}$)
featureless continuum (FC) should be added to the spectral base \citep{1978ApJ...223...56K} to account for the AGN
emission.
Values for $\alpha$ are between $0.5-2.0$ \citep{Osterbrock_2006}, with many authors working with large samples of galaxies assuming a general value of $\alpha=1.5$ \citep{2003ASPC..297..363S, 2003MNRAS.339..772R, 2004MNRAS.355..273C, Mallmann_2018}.
Since this work is focused on a single object, we propose a method to specifically fit the $\alpha$ index for the optical spectrum of NGC~2992. 
We extracted two integrated spectra: an inner one, hereafter $S(\lambda)$, defined as the sum of spectra within a circular region with radius of 0$\farcs$4, and an external one, $E(\lambda)$, in an annulus with inner radius of 1\arcsec and outer radius of 1$\farcs$4, both centred at the peak of continuum emission.
We model the $S(\lambda)$ spectrum as a combination of $E(\lambda)$ and a power law FC ($F_\lambda \propto\lambda^{- \alpha}$), both reddened by the same factor and normalised in the same wavelength, $\lambda_{0} = 5500$ \textup{\AA}, as follows:

\begin{multline}
\left( \frac{S_{m}(\lambda)}{S_{m}(\lambda_{0})}  \right) \\ = 
\left[ \left(1-\boldsymbol{p}  \right) 
\left( \frac{E(\lambda)}{E(\lambda_{0})}  \right)  +
\boldsymbol{p} 
\left ( \frac{ \lambda }{ \lambda_{0} } \right )^{- \boldsymbol{\alpha}} \right]
\times 10^{-0.4 \boldsymbol{\delta A_V} q_\lambda}    
\end{multline}

\vspace{0.5cm}

\noindent
where $S_m(\lambda)$ is the model spectrum, $p$ is the fraction contribution of the FC, $\delta A_{V}$ is the difference
in extinction between the $S(\lambda)$ and $E(\lambda)$ regions.
The values which minimise the residuals between $S(\lambda)$ and $S_m(\lambda)$ are $p$ = 0.23, $\delta A_V$ = 2.2 and $\alpha$ =1.7.

It should be noted we are over-simplifying by assuming that $S(\lambda)$ is the combination of $E(\lambda)$ and an FC, \textrm{i.e.}  the stellar spectrum of both are the same.
However, we argue this is just a method to get a data-based value for $\alpha$, instead of just assuming a fixed value
which is even less physically justified.
Also, no further interpretation will be given in this regard besides the adoption of this $\alpha$ value to produce the FC base element\footnote{
Further validation of this approximation will be presented throughout the paper, once the 
almost constant radial profiles shown in \autoref{fig:Radial} is taken into account, and the inferred stellar $A_{V}$ estimates, roughly consistent with those derived from the Balmer decrement (see also e.g. Fig. A.1 in \citet{Mingozzi_2019} for consistent results) will be presented in appendixes \ref{app1} and \ref{sec:Error}.
}.

\subsection{Spatially Resolved Stellar Populations Synthesis}
\label{sec:Resolved_SP}

In order to perform the spatially resolved stellar synthesis with enough signal to noise ratio ($S/N$) we employed the Voronoi tessellation technique \citep{2003MNRAS.342..345C} in our data cube, with a $S/N$ target set to 20.
Results from \cite{Cid_2005} show at this level {\sc starlight} is able to reliably recover input parameters.
Some spectra, however, failed to reach the desired $S/N$, even when combining several tens of spaxels, specifically those obscured by the galaxy's dust lane.
Henceforward we will just analyse and show the results for those voronoi zones which the combined-spectra has $S/N \geq 10$, relying again on \cite{Cid_2005} tests that show it to be the minimum value to the reliability of the code. \textsc{starlight} measures the $S/N$ as the ratio between the mean and the Root-Mean-Squared (RMS) flux in a user-selected region of the spectrum (which should be as featureless as possible), the range adopted here is a $\Delta\lambda$ = 20 \AA ~ wide window around the normalisation wavelength as measured from the combined-spectra of the Voronoi zone.

In order to be able to explore both the effects of the FC component and the change of model/library, we run the \textsc{starlight} code in the spaxels of the binned cube using three distinct spectral bases:

\begin{itemize}
    \item \textit{BC03}: Only stellar SSP components from the \citetalias{BC_2003}/STELIB models, see \autoref{table:SSPs} for ages and metallicities.
    \item  \textit{BC03+FC}: Same as \citetalias{BC_2003} stellar SSPs, with the addition of an FC component ($F_\lambda \propto\lambda^{- 1.7}$) to the spectral base.
    \item \textit{M11+FC}: Stellar SSP components from \citetalias{Maraston_2011}/MILES, see \autoref{table:SSPs} for ages and metallicities, in addition to the FC component.
\end{itemize}

\noindent
Since the effects of the presence/absence of FC component can be tested comparing \textit{BC03} with \textit{BC03+FC}, and the change of model/library is explored comparing \textit{BC03+FC} with \textit{M11+FC}, we did not find necessary to test the synthesis using \textit{M11} without FC, therefore we kept our analyse within these three configurations.

\begin{figure*}
    \centering
	\includegraphics[width=0.8\textwidth]{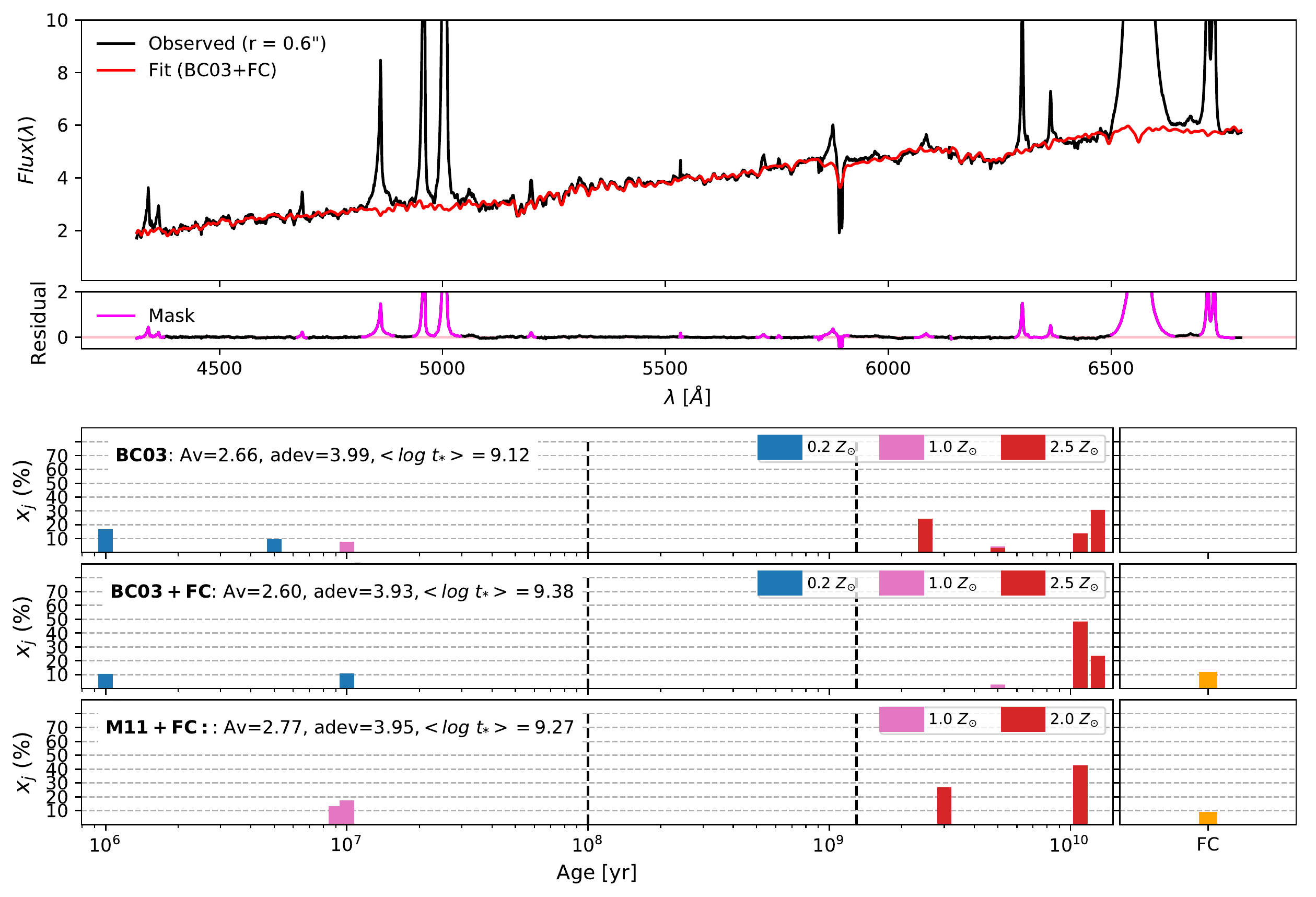}
    \caption{Example of stellar population synthesis fit. Top: Observed (black) and synthetic (red) spectra both in units of 10$^{-15}$\ergcms in a spaxel at radius equal to $0\farcs6$. Middle: Residual between the observed and synthetic spectra, masked emission lines are shown in purple. Bottom: Age/metallicity and FC decomposition for the three synthesis, \textit{BC03}, \textit{BC03+FC} and \textit{M11+FC}. The dashed lines show the division between $x_Y$, $x_I$ and $x_O$ SSPs.}
    \label{fig:Fit_Stellar}
\end{figure*}

The FC component is available for {\sc starlight} only in the spaxels of the inner 1.1 arcsec\footnote{The radius(r)$ = 0.0$ is set as the peak of the continuum emission.}, which equals $3\sigma$ of the Gaussian point spread function. 
The reason for not allowing the FC to be used throughout the field is that there is a high level of degeneracy between the FC and a young reddened population (further discussion in \citet{2010MNRAS.403..780C} and the next section), and also because there is no physical reason to have an FC emission outside the nucleus since it is emitted by the spatially unresolved accretion disk.
We will call the percentage contribution of the FC to total light of a spaxel as $x_{agn}$. During the synthesis all the emission lines were masked (see \autoref{fig:Fit_Stellar}), the mask is the same in all spaxels with exception to the central ones (r < 1.1\arcsec) where \hbeta and \nii+\halpha masks are broader. The sum of $x_Y$, $x_I$, $x_O$, $x_{agn}$ values is always 100 per cent for all spaxels.

The maps of the recovered properties from the three distinct bases are shown in \autoref{fig:BC03}, \autoref{fig:BC03+FC} and \autoref{fig:M11+FC} at the \refappendix{app1}. These figures show the maps for $x_Y$, $x_I$, $x_O$, $x_{agn}$, $A_V$, the light-weighted stellar mean age \meanAge defined as:

\begin{equation}
    \langle log \ t_{*} \rangle = \frac{\sum_{j=1}^{N_{*}} x_j log(t_j)}{\sum_{j=1}^{N_{*}} x_j}
\end{equation}

\noindent
where $x_j$ are the elements of the population vector $\Vec{x}$, i.e. the fraction contribution of each component to the total light, $t_j$ is the ages of $j$-th component, and $N_{*}$ is the number of stellar spectra in the base (SSPs). The figures in \refappendix{app1} also shows the light-weighted stellar mean metallicity $\meanZ$ for each reduced population vector component $x_Y$, $x_I$ and $x_O$, respectively $\meanZ_Y$, $\meanZ_I$ and $\meanZ_O$, defined as:

\begin{equation}
    \meanZ = \frac{\sum_{j=t_0}^{t_f} x_j z_j}{\sum_{j=t_0}^{t_f} x_j}
\end{equation}

\noindent 
where $t_0$ and $t_f$ are respectively the lower and upper age limit of the reduced population vector components as defined in the previous section.

The upper panel of \autoref{fig:Fit_Stellar} shows an example fit for a spaxel at $r = 0\farcs6$, the residuals between the observed and the fitted spectra and the emission line mask.
At the bottom panels the age/metallicity/FC decomposition for the three syntheses are shown as an example of the small differences resulting from the change of spectral base (see \autoref{sec:FC} and \autoref{sec:Model}). 

\begin{figure}
    \centering
	\includegraphics[width=0.95\columnwidth]{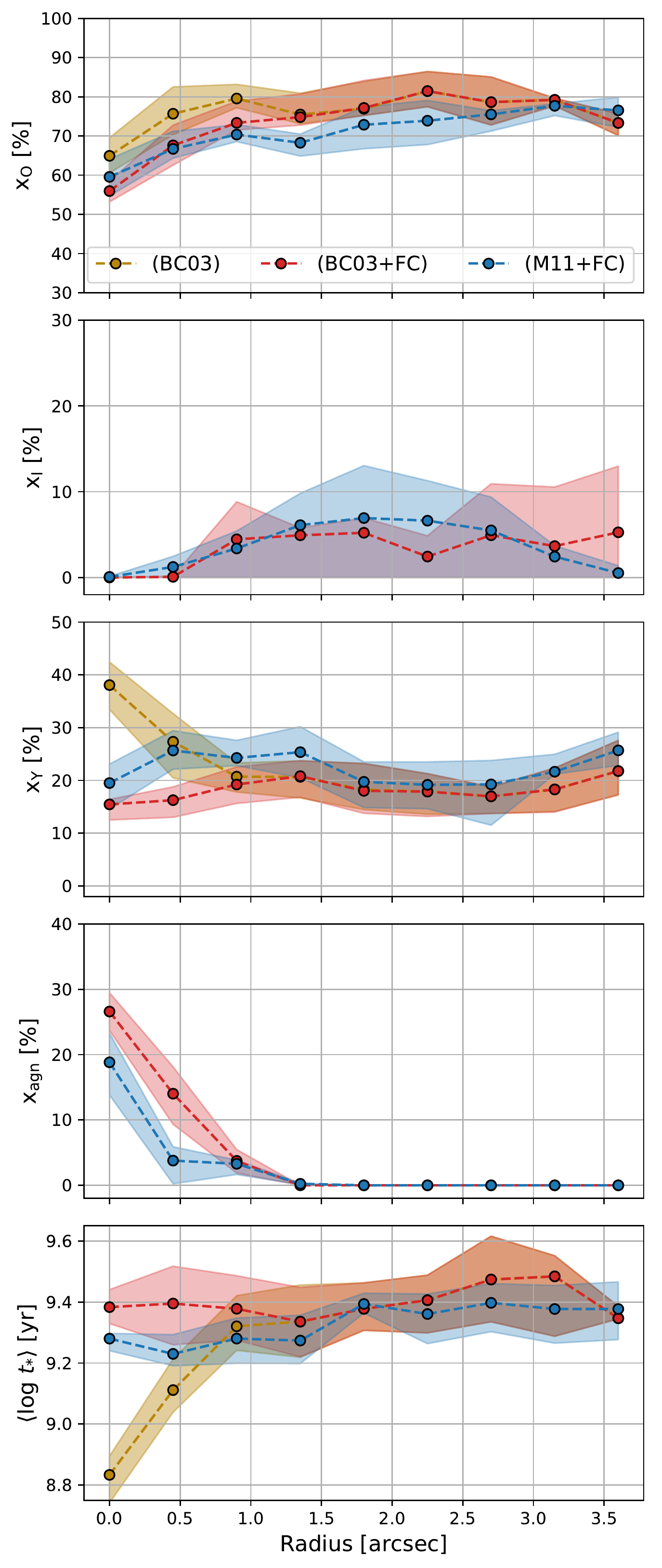}
    \caption{Radial profiles of the recovered properties for \textit{BC03} (gold), \textit{BC03+FC} (red)
    and \textit{M11+FC} (blue) in bins of $0\farcs45$ ($\sim$ 70 pc). From top to bottom: Per cent contribution
    to the total light from young ($x_O$), intermediate ($x_I$) and old ($x_Y$) SSPs, from the FC ($x_{agn}$) and the light-weighed mean stellar age (\meanAge). $x_{agn}$ is null for \textit{BC03} by definition. The $x_I$ values for  \textit{BC03} are not shown, however they are exactly
    the same as \textit{BC03+FC}.} 
    
    \label{fig:Radial}
\end{figure}

In \autoref{fig:Radial} we show the radial profiles of $x_Y$, $x_I$, $x_O$, $x_{agn}$ and \meanAge, in bins of 0$\farcs$45 ($\approx$ 70\,pc), with points representing the median, and the coloured region represents the 25 and 75 percentiles of the azimuthal distribution for the given radial bin. In \autoref{fig:Z} we show the values of \meanZ for each one of the bases integrated over the entire FoV.

In general terms the presence of a large contribution of old metal-rich SSPs and a smaller, but significant, contribution of young metal-poor stars seems to be independent of the presence of an FC component, or the change of model/library.
The detailed analyses of the differences caused by the distinct spectral bases available to the code will be better explored in the sections below.

\subsection{The FC Component Usage: Comparing BC03 and BC03+FC}
\label{sec:FC}

As shown by \autoref{fig:Radial}, the differences in the \textit{BC03} and \textit{BC03+FC} synthesis are seen only at the inner spaxels and are linked to the presence of the FC component in the spectral base.
For the external spaxels ($r > 1\farcs1$), where only \citetalias{BC_2003} stellar components are used in both synthesis, the recovered properties are identical, as one would expect.
In the external spaxels, the $x_O$ values range between 75 per cent and 85 per cent, with all spaxels having super-solar mean metallicity, $\meanZ_O > 1.5$\zsun, and the majority of them being close to the upper limit of the models, $ \meanZ_O  \approx 2.5$\zsun.
The $x_Y$ values are between 15 per cent and 25 per cent, with most spaxels having a sub-solar mean metallicity, $\meanZ_Y \leq 1.0$\zsun, with a considerable portion of them at the lower limit of the model, i.e.  $\meanZ_Y = 0.2$\zsun.
Some properties are identical in all runs even at the inner spaxels, $A_V$ is very high in the innermost spaxels ($A_V > 3$ mag) declining rapidly to  $A_V \approx 2.5$ mag and reaching 1.5 mag at the border spaxels ($A_V$ map shown at \refappendix{app1}).
Similarly, the contribution from intermediate SSPs, $x_I$, is null at most spaxels and reaches a maximum value between 10 and 20 per cent in very few regions.

Comparing the inner spaxels of \textit{BC03} and \textit{BC03+FC}, we see that the presence of an FC component in the base causes a decrease in $x_O$ by just a few per cent.
Larger differences, however, are seen in the contribution from young stars, especially where an FC component is not available (\textit{BC03}).
In this case, there is a sharp increase in $x_Y$ from a steady 20 per cent at $r > 1$\arcsec to more than 40 per cent at the centre.
On the other hand, the (\textit{BC03+FC}) base produces a roughly constant $x_Y$ at $r < 1.0\arcsec$, while $x_{agn}$ increases from $\sim$0 per cent at 1.0\arcsec to $\sim$20 per cent in the innermost spaxels. 
This effect is caused by the above-mentioned degeneracy between an FC spectrum and a young-reddened SSP \citep{2004MNRAS.355..273C,2010MNRAS.403..780C}.
As shown by \citet{2017A&A...604A..99C}, fitting a spectrum that contains an AGN power law-like continuum with only stellar components causes the solution to have an excess of young populations, which leads to a drop in the \meanAge (see the bottom panel of \autoref{fig:Radial}).
Although such degeneracy can cast some doubts in the exact values of the $x_Y$ and $x_{agn}$ decomposition in \textit{BC03+FC}, the fact that the solution at the centre has $x_Y$ values compatible with the ones of the external regions of the FoV, and that the FC component contribution follows a radial profile compatible with an unresolved source (i.e increasing from $x_{agn} \approx$ 0 per cent to $x_{agn} \approx$ 20 per cent at the continuum peak) is strong evidence in favour of the existence of an FC emission in NGC~2992.

\subsection{Decomposition Stability: Comparing BC03/STELIB and M11/MILES}
\label{sec:Model}

In this section we discuss differences in the recovered properties between \textit{BC03+FC} and \textit{M11+FC} bases and their derivations.
We do not intend to make a detailed analysis of the models and its derivations, nor to suggest one is a better choice than the other for use with this type of data.
Such discussions can be found in other studies, e.g. \citet{Baldwin_2018, Ge_2019}.
Instead, we merely want to test the stability of the inferred stellar properties with changes in models.

Besides the distinct stellar libraries used in these model libraries, they also employ different evolutionary prescriptions  \citep[see][for a review on stellar evolutionary models]{Conroy_2013}.
\citetalias{BC_2003} is constructed using the isochrone synthesis approach, while \citetalias{Maraston_2011} uses the fuel consumption theory. 
The treatment of the thermally pulsing asymptotic giant branch (TP-AGB) phase is also a major topic of disagreement between different models.
For example, \citet{Dametto_2019} found a relative excess of young populations and lack of intermediate-age components when Near-Infrared (NIR) stellar populations synthesis are performed with \citetalias{BC_2003} SSPs,
favouring results achieved with the \citetalias{Maraston_2011} models.
Nevertheless, these effects are much more important when modelling stellar populations in the NIR, being significantly less relevant in the optical.

Both $x_{agn}$ and $A_V$ are very similar to those obtained with \textit{BC03+FC}, but the population vectors have notable differences.
For instance, $x_O$ values for \textit{M11+FC} are systematically lower by $\sim$5 to 10 per cent than those recovered with \textit{BC03+FC}, while at the same time both $x_Y$ and $x_I$ are conversely higher for \textit{M11+FC}.
Given the increase of both $x_Y$ and $x_I$ and the decrease in $x_O$, the light-weighted mean stellar age \meanAge for \textit{M11+FC} is systematically lower than for \textit{BC03+FC} by $\sim$ 0.5 to 0.1 dex, as show in the bottom panel of \autoref{fig:Radial}.

\begin{figure}
    \centering
	\includegraphics[width=\columnwidth]{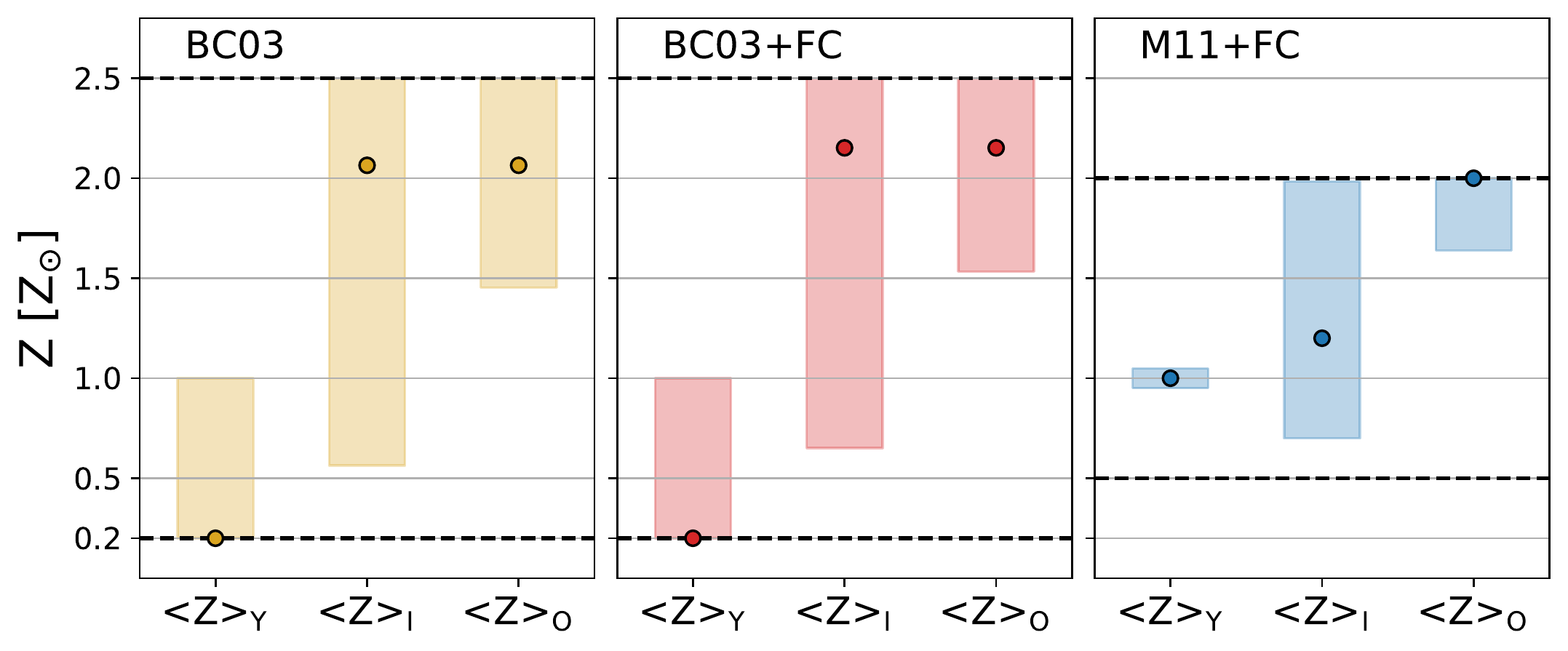}
    \caption{Values of \meanZ$_Y$, \meanZ$_I$ and \meanZ$_O$ integrated over the entire FoV for each base, from left to right, respectively, \textit{BC03}, \textit{BC03+FC} and \textit{M11+FC}. Points are the median values, and the coloureds bars are the 25 and 75 percentiles. The dashed black lines show the metallicity limit for each of the models.} 
    \label{fig:Z}
\end{figure}

Although there are small differences in age decomposition, the metallicity general pattern seems to be maintained.
As can be seen in \autoref{fig:Z}, the old populations have a super-solar mean metallicity, close to the upper limit of the model ($ 1.5$\zsun $< \meanZ_O \ \leq 2.0 $\zsun) and the young populations have a solar to sub-solar mean metallicity ($\meanZ_Y \ \leq 1.0$ \zsun), as shown in the middle panels of \autoref{fig:Z}.
However, while the values of $\meanZ_Y$ in \textit{BC03+FC} are mostly sub-solar ($\meanZ_Y < 1.0$ \zsun) the values in \textit{M11+FC} are all solar ($\meanZ_Y = 1.0$ \zsun).
This discrepancy is probably caused by the fact that the MILES library does not contain SSPs with a sub-solar metallicity younger than 55 Myr (see \autoref{table:SSPs}).
For instance, in the example of \autoref{fig:Fit_Stellar}, while \textit{BC03} and \textit{BC03+FC} recover its young populations as a combination of three SSPs with ages of 1 Myr, 5 Myr and 10 Myr, the two first with sub-solar metallicity ($Z = 0.2$\zsun) and the later with solar metallicity ($Z = 1.0$\zsun), the \textit{M11+FC} recovers it as both 9 Myr and 10 yr solar metallicity SSPs, and the fitting code would not be able to fit such a young component with lower metallicity precisely because MILES library does not provide an SSPs with $Z = 0.5$ \zsun younger than 55 Myr.
Therefore, we can see these $\meanZ_Y = 1.0$\zsun values in the synthesis using \textit{M11+FC} base as an upper limit for the young population metallicity.

As mentioned, above, we do not claim any of the three syntheses to be the one that better describes the data.
In fact, there are no significant differences in the \textit{adev} parameters among the bases: \textit{adev} values range from 3 per cent at the inner spaxels up to 8 per cent at larger radius.
Recovered properties produced with the distinct models present the same general trend (see \autoref{sec:Resolved_SP}). However, we argue the bases which include an FC component are more physically justified. Both the large decrease in \meanAge in the inner spaxels, when this component is ignored, and the fact that the FC components fluxes have a spatial profile characteristic of a point-like source (FWHM $\approx$ $0\farcs8$, see \autoref{fig:BLR}) are strong evidence of the existence of an FC like emission in NGC 2992.
It is worth mentioning even though we are not discussing results for the stellar kinematics, we have made sure all the three decompositions were kinematically consistent with each other, and no non-physical value was fitted to $v_*$, $\sigma_*$ parameters, which would cause an incorrect subtraction of the continuum used in emission line fitting procedures of the next sections.

\section{GAS KINEMATICS}
\label{sec:kinematics}

Emission line structure of the central regions of AGN is very complex, often possessing multiple velocity components along the line of sight \citep[e.g.][]{Villar-M_2011, Veilleux_2013,McElroy_2015}.
For the type-1 and intermediate-type AGNs, the presence of both narrow and broad emission lines are seen; these emission lines are believed to come from two distinct parts of the AGN. 
The first is named the "Broad Line Region" (BLR).
Its typical size, as deduced from the broad line variability, is 10-100 light-days, therefore it is spatially unresolved by current observations, meaning that its spatial profile is the same as the point spread function (PSF) of the observation.
BLR emission is characterised by very broad (1000\kms < FWHM < 10000\kms) Balmer lines and the absence of forbidden lines \citep{Peterson_1997}. 
In most cases, the line profiles strongly deviate from a Gaussian profile, e.g. NGC~1097 \citep{Schimoia_2015}.
The second region is the "Narrow Line Region" (NLR) which is much larger than the BLR since no clear variation of the narrow emission lines is observed in objects undergoing large variations of the AGN's featureless continuum on timescales shorter than several years.
The NLR is resolved by ground-based observations showing dimensions of 100-1000 pc in most nearby AGNs and up to several kpc's in higher redshift quasars \citep{Bennert_2006, Sun_2017,Storchi-Bergmann_18}.
The NLR emission is characterised by both permitted and forbidden lines with lower velocity dispersion (FWHM < 1000\kms), but still broader than emission lines of normal star forming galaxies.
Even the NLR emission lines are some times not well fitted by a simple Gaussian profile, given the presence of multiple kinematical components in the line of sight, e.g. gaseous disk rotation, gas outflows and inflows \citep[e.g.][]{Zakamska_2016,Schnorr_2017}.

\subsection{Emission line fitting}
\label{sec:fit_EM}

In order to fit the emission lines in NGC~2992 we use the \textsc{IFSCube} package\footnote{http://github.com/danielrd6/ifscube} \citep{ifscube}.
\textsc{IFSCube} is a python based software package designed to perform analysis tasks in data cubes of integral field spectroscopy, mainly focused on the fitting of spectral features, such as atomic lines.
It allows for multiple combinations of Gaussian and Gauss-Hermite functions, with or without constraints to its parameters.
Stellar continuum subtraction, as well as pseudo continuum fitting if necessary, are performed internally, and in this case, we used the residual spectra from the stellar population synthesis performed with {\sc starlight}.
Initial guesses for the parameters are set for the first spaxel, usually, the one with the highest $S/N$, and all subsequent spectra use the output of neighbouring successful fits as the initial guess.

When dealing with complex emission, the standard approach is to model the line profile with a combination of Gaussian functions.
Visual inspection of our IFU shows distinct velocity components, evidenced by large asymmetries in the line profiles, are present in over one-third of the spectra in the FoV.
As a result, we decided to model these components separately and to investigate whether they possessed any meaningful physical information, following previous works \citep[e.g.][]{Villar-M_2011, McElroy_2015, Fischer_2018}.

For every fitted emission line, we determined if the multi-component model was statistically justified, and not a better fit purely by virtue of the extra model parameters, by performing a series of f-tests.
The f-test is a standard statistical test to gauge whether a higher order model is preferable to a simpler model when fitting a particular data set.
We set the false rejection probability for the lower order model to 10$^{-5}$.
The higher this threshold is set, the harder it is to justify a more complex model and the more likely it is to be over-fitting noise.
The reader interested in more detailed analyses and a step-by-step deduction of the application of f-test in astrophysics emission line fitting is referred to \citet{Freund_1992} and statistical references therein.

As an Intermediate-type Seyfert nucleus, NGC~2992 presents both BLR and NLR emission.
The BLR emission lines (\halpha and \hbeta) are blended with some NLR emission lines (\halpha, $\nii\lambda6548$, $\nii\lambda6583$ and \hbeta), leading to degeneracies if both NLR and BLR emission lines are fitted, spaxel by spaxel, at the same time with all parameters free.
Fortunately, the fact that the BLR is not resolved, i.e. its kinematics and its Balmer decrement (\halpha/\hbeta) can not vary spatially, can be used to decrease this degeneracy.
Our chosen methodology consists on first fitting an integrated spectrum to fix the BLR emission line profile (kinematics and Balmer decrement), and later using this BLR spectrum as a fixed component, except for the flux, on the spatially resolved NLR fits.

\begin{figure*}
    \centering
	\includegraphics[width=0.9\textwidth]{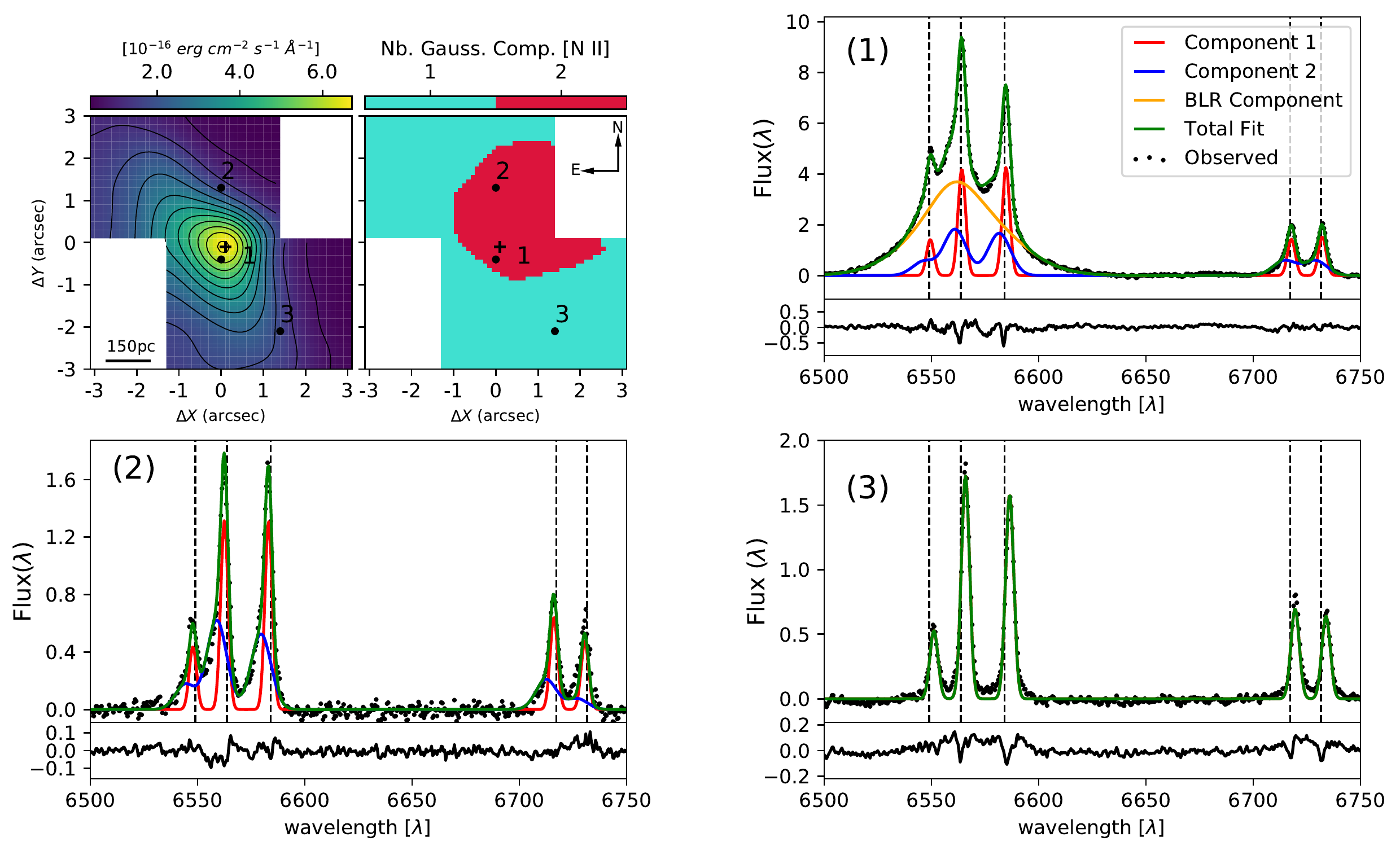}
    \caption{Top left: Map of the continuum emission at $\lambda$ = 5500 $\textup{\AA}$ and map of the preferred number of Gaussian components to the \nii emission-lines. Top right and bottom:  Examples of the \halpha, $\nii\lambda6548,\lambda6583$ and $\sii\lambda6716,\lambda6930$ emission lines fitting and their residual in distinctly regions of the FoV marked in the maps. The vertical axis units in the panels showing the emission-line profiles are 10$^{-15}$\ergcms and the dashed vertical lines show rest-frame wavelength of the lines.}
    \label{fig:NII}
\end{figure*}

We extracted a spectrum centred at the continuum peak, using a virtual aperture of $3\sigma = 1\farcs1$ ($\sigma$ refers to the Gaussian point spread function).
In this spectrum we fitted $\nii\lambda6548$, $\nii\lambda6583$, \halpha (both BLR and NLR) and \hbeta (both BLR and NLR), adding Gaussian components to the emission lines until the addition of a new component does not increase the fitting quality, as measured by the f-test described above.
The final fit has shown that two Gaussians are necessary to fit each one of the BLR Balmer emission lines.
This resulting profile, meaning the superposition of two Gaussian functions, hereafter called as BLR Component, can be seen in panel (1) of \autoref{fig:NII}, and is characterised by a FWHM$_{\rm BLR} = 2010$\kms and a Balmer decrement of (\halpha/\hbeta)$_{\rm BLR} = 24.2$.
These values are in considerable agreement with the ones fitted by \citet{Schnorr_2016}, FWHM$_{\rm BLR} = 2192$\kms and (\halpha/\hbeta)$_{\rm BLR} = 19.2$, considering these authors used three Gaussian functions to model the profile and the BLR of NGC~2992 has been shown to vary in a few years period \citep{Trippe_2008}.

In order to resolve the kinematics of the NLR and obtain maps of emission line ratios, we fitted the following eight strongest emission lines present in the data cube: \hbeta, $\oiii\lambda4959,\lambda5007$, \halpha, $\nii\lambda6548,\lambda6583$ and $\sii\lambda6716,\lambda6731$.
Each emission line was modelled with a set of Gaussian functions.
The central wavelength of the Gaussian functions was allowed to vary by up to $\pm$400 \kms and the velocity dispersion up to 400 \kms (FWHM $\approx 1000$ \kms).
We added one more set of Gaussians to the fit, meaning that each emission line is modelled by up to two Gaussian components.
To reduce the number of free parameters, each set of Gaussians was required to have the same velocity and velocity dispersion (meaning their relative wavelengths were fixed), while their fluxes were allowed to vary.
The reason for this is each component represents a kinematically distinct part of the gas, meaning that they may have different ionisation states determined by their relative fluxes.
This means that, for example, the single-component set of Gaussians fit to the eight emission lines instead of having twenty four (eight lines $\times$ three parameters) free parameters, only had nine (a single velocity, $v$, and velocity dispersion, $\sigma$, amplitudes of \halpha, \hbeta, the two \sii and the stronger \nii and \oiii, while the amplitudes of the weaker \nii and \oiii were set fixed as 1/3 of the stronger ones, as given by their Quantum Mechanics' probabilities (hereafter we call $\oiii\lambda5007$ as \oiii, $\nii\lambda6583$ as \nii, and the $\sii\lambda6716,\lambda6731$ as \sii).
As described in the previous paragraph, the BLR Component was also included in the fit with only the amplitude as free parameter.

To estimate the uncertainties in the parameters, we performed Monte Carlo simulations, in which Gaussian noise was added to the observed spectrum.
One hundred iterations were performed and the estimated uncertainty in each parameter was derived from the standard deviation in all the iterations. The uncertainty maps are presented in \refappendix{sec:Error}.

\begin{figure*}
    \centering
	\includegraphics[width=0.9\textwidth]{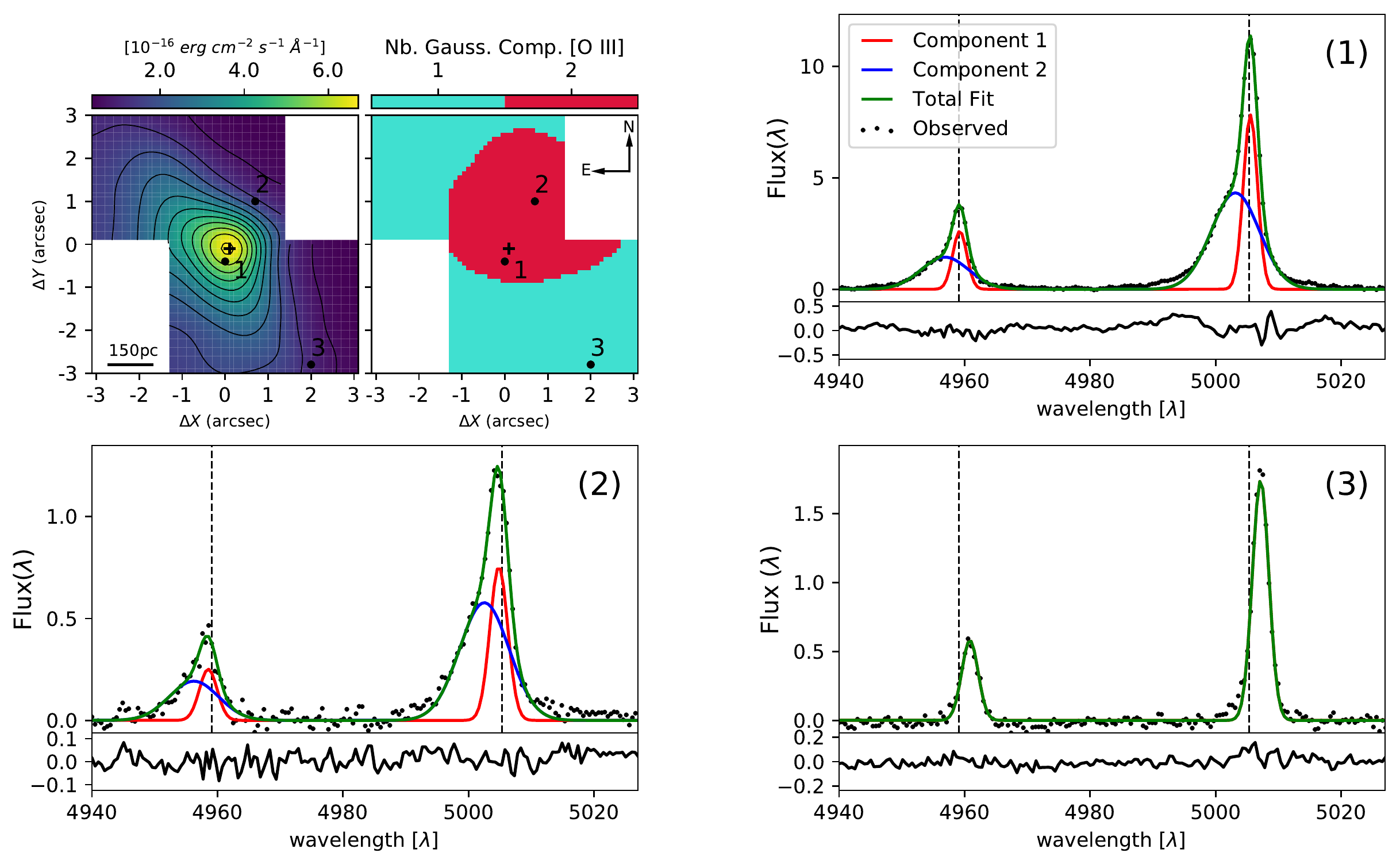}
    \caption{Same as \autoref{fig:NII} but for \oiii emission lines.}
    \label{fig:OIII}
\end{figure*}

We then applied the f-test to each one of the NLR emission lines, spaxel by spaxel, in order to assess if the simple model (single-Gaussian) is enough to fit the data or the more complex model (two Gaussians) is required.
The f-test results in spatially contiguous regions of the required number of components for each emission line. The maps for \nii and \oiii emission lines are shown in \autoref{fig:NII} and \autoref{fig:OIII}. The figures also show spectral and emission line fits in selected regions of the FoV.

The maps are not exactly equal for all emission lines, the size of the region in which two components are needed depends on the line S/N\footnote{
    Because the f-test detection is sensible to how much the second component, the more complex model, stands out from the noise.
    Therefore in low S/N lines, it is less likely the f-test will require a complex model.},
therefore the size is maximum for \oiii (\autoref{fig:OIII}) and \halpha (\autoref{fig:Outflow}), a little bit smaller for \nii (\autoref{fig:NII}) and the \sii doublets (not shown) and much smaller for the low S/N line, \hbeta (not shown).
However, the general pattern is the same for all emission lines: the data are well fitted by a single-Gaussian in most of the FoV, but in a region, from the nucleus to the end of the FoV in the northwest (NW) direction two Gaussians are needed.

As can be seen in both figures, for those spaxels in which the emission lines are well behaved and symmetric, a single-Gaussian function, which we will call Component 1, is enough to fit the data.
However, in the spaxels where two components were statistically significant, the profiles show an asymmetry in the form of a blueshifted wing (see panels (1) and (2) in \autoref{fig:NII} and \autoref{fig:OIII}) requiring the addition of a second Gaussian, which we will call Component 2.

\begin{figure}
    \centering
	\includegraphics[width=\columnwidth]{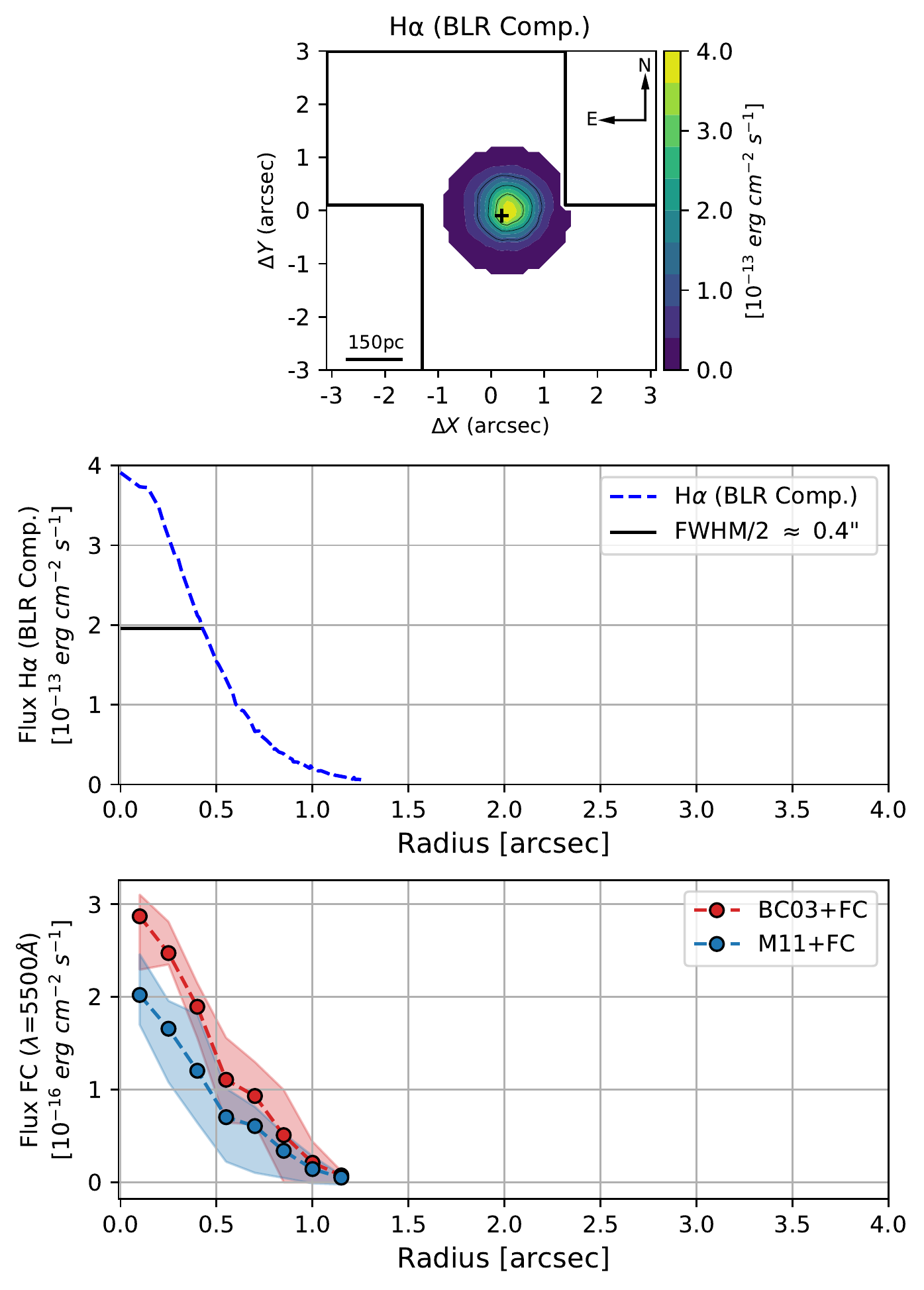}
    \caption{Top: Spatial distribution of the BLR Component. Middle: Radial projection of the BLR Component profile, vertical axis units of  10$^{-13}$\fluxline. Both show the point-like source emission of the BLR, with a FWHM $\approx 0\farcs8$ given by the spatial resolution of the observation. Bottom: Radial projection of the FC components fluxes (see \autoref{sec:stellar}), also presenting a point-like source emission.}
    \label{fig:BLR}
\end{figure}

In \autoref{fig:BLR} we show the spatial and radial profile of the BLR Component.
As expected for an unresolved structure, this component is detected just in the innermost spaxels, with a spatial distribution of a point-like source, equal to that of the stars in the acquisition image (FWHM $\approx 0\farcs8$, see \autoref{sec:observations}); at radii larger than $3\sigma$ (1.1 arcsec) it is almost undetectable.

\subsection{Rotational and Non-Rotational Components in the NLR}

\begin{figure}
    \centering
	\includegraphics[width=\columnwidth]{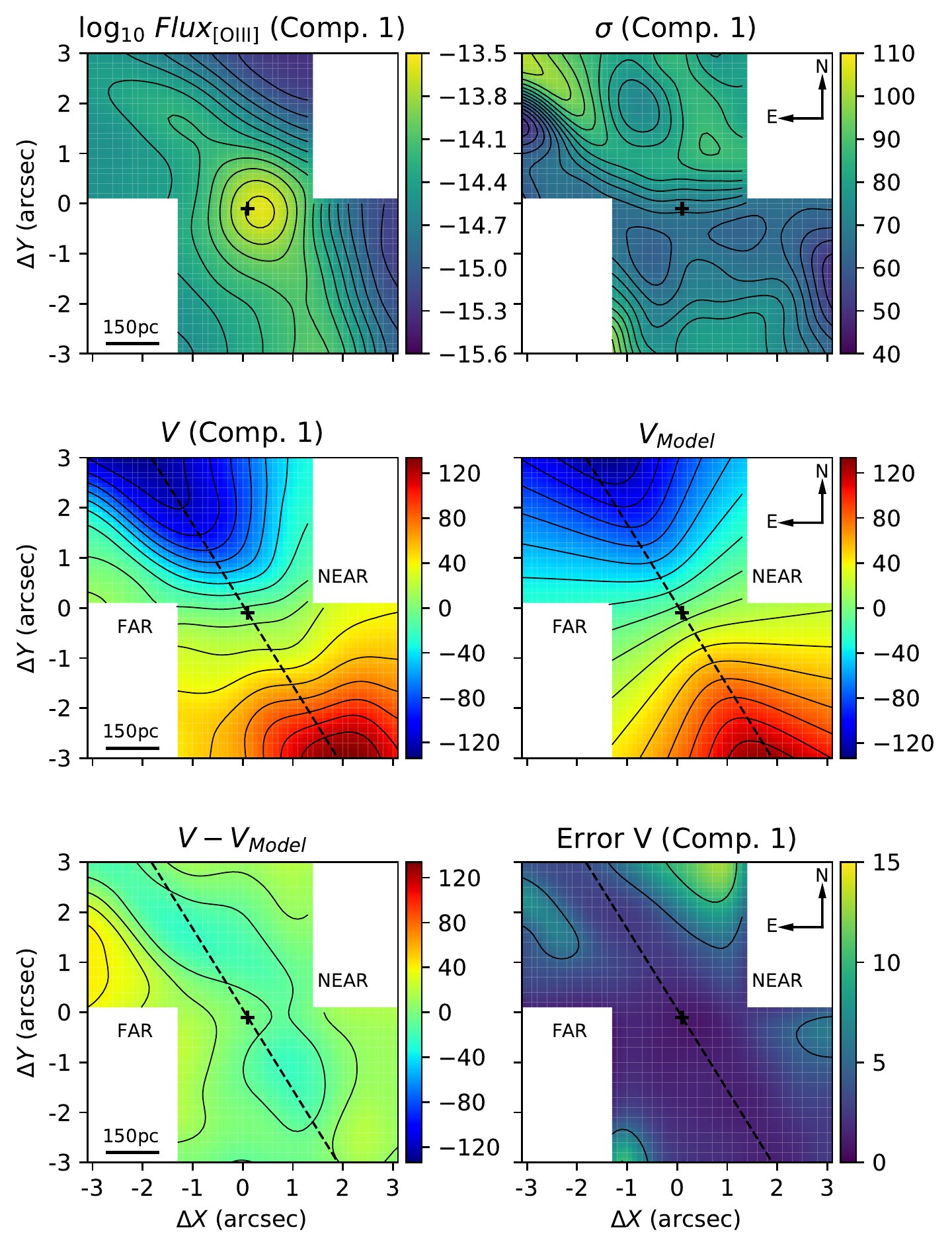}
    \caption{Component 1 properties. Top left: Log$_{10}$ Flux
    \oiii$\lambda$5007 (\fluxline). Top right: Velocity dispersion $\sigma$
    (\kms). Middle left: Radial velocity $v$ (\kms). Middle right: Modelled radial
    velocity (\kms), from the \citet{Bertola_1991} rotation model fitting. Bottom
    left: Residual between the measured and the modelled velocity field.
    Bottom right: Uncertainties in the radial velocity $v$ (\kms). Black cross is the continuum peak and the dashed black line is the fitted major axis PA.}
    \label{fig:Comp1}
\end{figure}

Let us now discuss the spatially resolved properties of the NLR Components 1 and 2 and their possible physical interpretation. 
Component 1 is present in the entire FoV.
In the top panels of \autoref{fig:Comp1} we show, from left to right, \oiii flux, velocity dispersion ($\sigma$), already corrected by instrumental dispersion. In the middle left panel, we present the radial velocity ($v$).
We show the \oiii flux as an example of the morphology of this component, which is similar to the continuum morphology, with the peak of the emission shifted by only $\sim 0\farcs2$ to southwest (SW), with a steep decrease (up to 2 $dex$) towards the dust lane region.
The $\sigma$ values range from 50-100\kms and show a very disturbed pattern.
The $v$ map shows values from -130 to 120\kms and although it presents some disturbance, a rotation pattern is clearly discernible, with the SW portion receding and the northeastern (NE) approaching.
We have considered the west side of the galaxy is the near one, following the literature \citep[e.g.][]{Marquez_1998,Veilleux_2001}; this explains why the emission is seen mainly to the east and less to the west where it is attenuated by the dust lane in the line-of-sight.

In order to quantify how much the velocity field deviates from a pure rotation and to derive both the major axis PA and the inclination of disk we fit the \cite{Bertola_1991} model. The model assumes a spherical potential with pure circular orbits, in which the observed radial velocity at a position ($R$, $\psi$) in the plane of the sky given by the relation:

\begin{multline}
    v(R, \psi) =  v_{\rm sys} +\\ 
    \frac{AR \cos(\psi - \psi_0) \sin \theta \cos^p i}{
        \left\{  R^2 \left[ \sin^2 (\psi - \psi_0) + \cos^2 (\psi - \psi_0) \right] 
        + c_0^2 cos^2 i \right\}^{p/2}
    }
\end{multline}

\noindent
where $v_{sys}$ is the systemic velocity, $i$ the inclination of the disk (with $i$ = 0 for a face-on disk), $A$ is the amplitude of the curve, c$_0$ is the concentration parameter and $\psi_0$ the line of nodes PA. 
The parameter $p$ defines the form of the rotation curve, varying in the range between 1 (logarithm potential) and 1.5 (Keplerian potential).
To avoid the degeneracy between the parameters, we restricted the disk inclination to have values of $i$\,=\,70\,$\pm$\,10 \citep[based on][photometric analyses]{Duc_2000} and we assumed the kinematical centre to be co-spatial with the peak of the continuum emission. 
A Levenberg-Marquardt least-squares minimisation was performed to determine the best fitting parameters and the uncertainties were obtained by making one hundred Monte Carlo iterations.
We show the best parameters and their uncertainties in \autoref{table:Bertola}.

\defcitealias{Veilleux_2001}{V01}
\defcitealias{Marquez_1998}{M98}

\begin{table}
\caption{Best Fitting Parameters \citet{Bertola_1991} Model.}
\centering
\begin{tabular}{|c|c|}
\hline
\hline
Parameters & Values  \\\hline
$v_{sys}$ & 2345 $\pm$ 5 [\kms]            \\
$A$           & 132 $\pm$ 4 [\kms]             \\
$p$           & 1$^a$                              \\
$c_0$        & 3.5 $\pm$ 0.1 [arcsec]                                    \\
$\psi_0$    &  $32^\circ\pm 1^\circ $  \\
$i$         & $70^\circ \pm 1^\circ  $ \\
\hline
\end{tabular}
\\
\footnotesize{$^a$ In the model the parameters $p$ is allowed to vary between 1 and 1.5, the fitted value was equals to 1 in all the Monte Carlo iterations.}
\label{table:Bertola}
\end{table}

The PA of the line of nodes was measured as 32$\pm$1$^\circ$, in agreement with previous values from  the literature: PA = 32$^\circ$ \citep[][hereafter V01]{Veilleux_2001}, PA= 30$^\circ$ \citep[][hereafter M98]{Marquez_1998}. V$_{sys}$ and \textit{i} are also in reasonable agreement with other studies of the galaxy: \citetalias{Veilleux_2001} found, respectively, $2335\pm20$\kms and 68$^\circ$, while \citetalias{Marquez_1998} found 2330\kms and 70$^\circ$. The velocity field amplitude (A = 132$\pm$4\kms), though, is much lower than the ones found by both \citetalias{Veilleux_2001} and \citetalias{Marquez_1998} (225$\pm$20\kms and 250\kms).
This discrepancy is probably due to the fact that both have used long-slit data in their studies, therefore being able to map the velocity field up to several kpc. Thus having a better constraint in the behaviour of the field at a larger radius, while our data can only recover the internal portion of the velocity field. 

\begin{figure}
    \centering
	\includegraphics[width=\columnwidth]{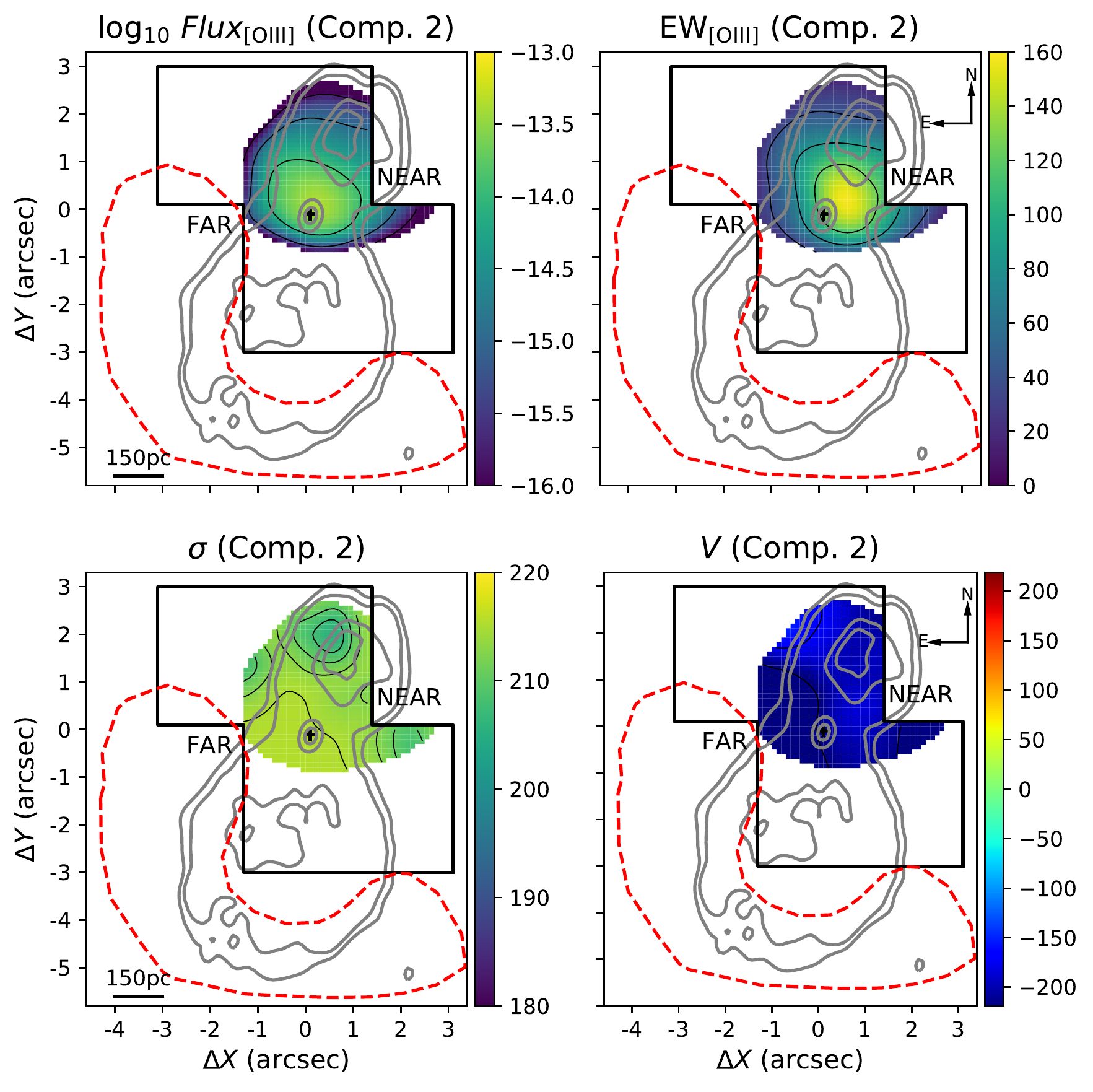}
    \caption{Component 2 properties. Top left: Log$_{10}$ Flux
    \oiii$\lambda$5007 (\fluxline). Top right: Equivalent Width of
    \oiii$\lambda$5007 ($\textup{\AA}$). Bottom left: Velocity dispersion, $\sigma$
    (\kms). Bottom right: Radial velocity, $v$ (\kms). Solid black lines represent our FoV. Black cross is the
    continuum peak. Grey contours are the 6cm radio emission
    \citep{Ulvestad_1984}. Red dashed line is the contour of the redshifted
    structure found by \citet{Garcia_lorenzo_01}, at the farthest part of the radio emission major loop, taken from Fig. 9 of their paper.}
    \label{fig:Comp2}
\end{figure}

The modelled velocity field is shown in the middle right panel of \autoref{fig:Comp1}, and the residuals between the observed field and model are shown in the bottom left panel of the same figure.
The uncertainty map for the Comp. 1 $v$ field, as measured by the Monte Carlo simulations described in the \autoref{sec:fit_EM}, is shown in the bottom right of \autoref{fig:Comp1}.
While most of the field is well represented by the disk rotation model ($\left | v-v_{Model} \right | < 20$\kms), there is one region at the eastern border with residuals up to 40\kms. 
Since the maximum uncertainty in $v$ is 15\kms, this deviation from pure rotation is probably real.
The same seems to be true for the $\sigma$ field, shown in \autoref{fig:Comp2}.
The uncertainties in the Component 1 $\sigma$ are of $\sim$5\kms in the central region, and up to 15\kms at the borders, as shown in \autoref{fig:fit_err}. 
Therefore, the lack of structure in the $\sigma$ map is, in fact, real and not due to observational artefacts\footnote{
    There is no clear evidence of the "Beam Smearing" effect too, given that there is no increase in velocity dispersion values of Comp.1 in the central regions, where the gradient of the velocity field is maximum.
    The high values of velocity dispersion in Comp. 2, also cannot be purely due to this effect, because it is not co-spatial with the galactic centre and shows very blueshifited velocities from -250 to -200\kms, as can be seen \autoref{fig:Comp2}. Furthermore when $\sigma$ of both Comp. 2 and Comp. 1 are taken into account their values are in agreement with the ones measured from Br$\gamma$ IR lines by \citet[][]{Friedrich_2010}.}.

These types of disturbances, in the both $v$ and $\sigma$, are predicted by simulations \citep{Kronberger_2007,Bois_2011} and observed \citep{Torres-Flores_2014,Hung_2016,Bloom_2018} in galaxies which are, or recently were, undergoing merging processes, as is the case of NGC~2992, and are attributed to perturbations in the gravitational potential due to tidal forces.
In Paper\,II we will investigate whether these disturbances are also present in the velocity field of NGC~2993.

\begin{figure}
    \centering
	\includegraphics[width=\columnwidth]{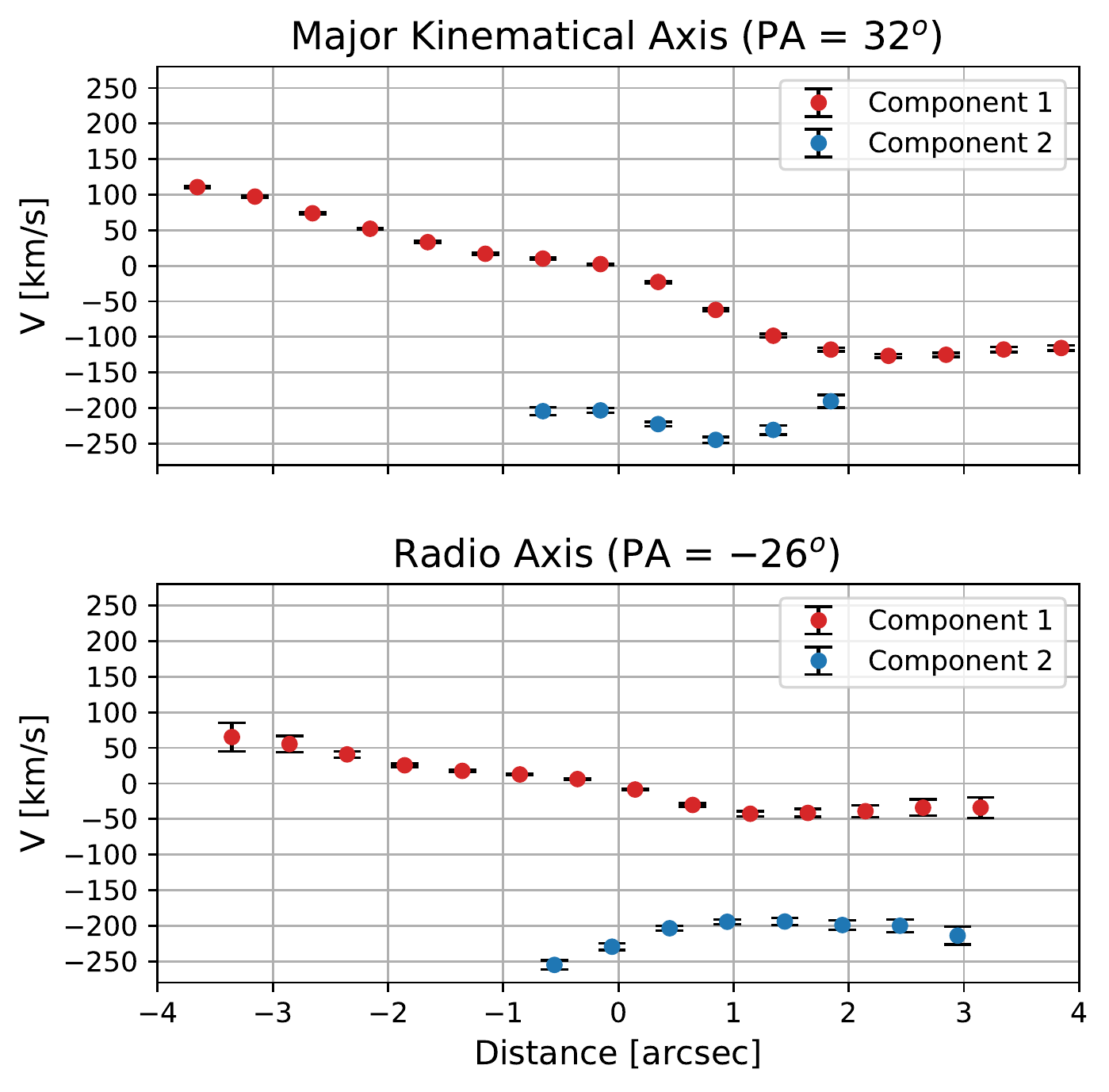}
    \caption{Position-Velocity plot in virtual slits with 0\farcs4 width. In the top panel the slit is at the PA of the major kinematical axis, in the bottom the slit is at the PA of 8-shaped radio emission major axis. Points are the measured velocity values, and the error bars are the uncertainties in the velocities. Red points are the velocities of Component 1 and blue points the Component 2 ones.}
    \label{fig:PV}
\end{figure}

The fainter component of the NLR (Component 2) is detected in a smaller region, 
originating near the continuum peak and extending towards the NW.
For the brightest line, $\oiii\lambda5007$, the emission extends up to $\sim$ 3\arcsec, as shown by both the flux map, top left panel of \autoref{fig:Comp2} and the P-V plot in \autoref{fig:PV}.
In \autoref{fig:Comp2}, we show the maps of equivalent width (EW) of \oiii, $\sigma$ and $v$ for the Component 2.
The $v$ map shows only blueshifted velocities, with values ranging from -250 to -200\kms.
The velocity dispersion is higher than that of Comp. 1, ranging from 200 to 215\kms.
We also show in all the panels, the contours of the 8-shaped region detected in the 6~cm radio emission (in grey)  
in which the smaller loop seems to be co-spatial with Component 2.

The position-velocity plot along the major kinematical axis
(PA = 32$^\circ$) and the major axis of the Radio emission (PA = -26$^\circ$) of NGC~2992 are shown in
\autoref{fig:PV}. It is clear that while Comp. 1 has a rotational pattern, Comp. 2 deviates 
considerably from this pattern. Considering: 

\begin{itemize}
    \item Component 2 is not following the galaxy's rotation;
    \item It is co-spatial with the radio emission;
    \item It has higher velocity dispersion values than those of Comp. 1, $\sigma >$ 200\kms;
    \item It has blueshifted velocities in the near side of the galaxy;
\end{itemize}

\noindent
we are led to the conclusion that it is tracing a radio-jet driven gas outflow since all these are well-known characteristics of this type of phenomena.
In fact, this ionised gas outflow was already reported by other IFU studies, e.g. \citet[][]{Garcia_lorenzo_01}, \citet{Friedrich_2010} and \citet[][]{Mingozzi_2019}. Particularly, \citet[][]{Garcia_lorenzo_01} interpreted this blueshifted structure as the interaction of the minor loop of the radio-jet with the interstellar medium (ISM).
\citet{Garcia_lorenzo_01} also found a similar, however redshifted, structure in the far side of the galaxy, co-spatial with the farthest part of the radio emission major loop, which is mostly not visible in our FoV.
We have extracted the contours of this structure from Figure 9 of their paper and show it as the red dashed line in \autoref{fig:Comp2}.
This finding supports the scenario in which a bipolar over-pressure relativistic plasma is expanding, accelerating and compressing the ISM gas in its path \citep{Chapman_2000}. 
Interestingly, the only portion of this structure which falls within our FoV is co-spatial with the largest velocity residuals seen in figure \autoref{fig:Comp1}.
Although our statistical analysis does not indicate the necessity of a separate component, the large residuals clearly indicate a perturbation in the velocity field.

A very similar geometry was found by \citet{Riffel_2011} in the galaxy Mrk 1157, in which non-rotational blueshifted and redshifted structures are present and co-spatial to the two loops of a very similar nuclear 8-shaped radio emission.
The authors also interpreted these structures as the effect of the expanding radio-emitting gas into the ionised ISM. 

\subsection{Outflow Properties}
\label{sec:Out_Prop}

If we consider Component 2 as ionised outflowing gas, we can measure its properties such as mass, mass outflow rate and kinetic power. We followed the prescriptions described in \citet{Lena_2015}, in which a biconical outflow geometry is assumed, even though in our case we can see only one of the cones.

The mass outflow rate can be estimated as the ratio between the mass of the outflowing gas and the dynamical time (time for the gas to reach the present distance from the nucleus), M$_{g}$/t$_{\rm d}$. The mass of the gas in outflow (Component 2) is estimated as:

\begin{equation} \label{eq:Mg}
M_{g} = n_e m_p V f
\end{equation}

\noindent
where $m_p$ is the proton mass, $n_e$ and $V$ are the electron density and the volume of the region
where the component is detected, $f$ is the filling factor.
The filling factor and volume are eliminated by combining \autoref{eq:Mg} with the following expression for the \halpha luminosity:

\begin{equation} \label{eq:Lha}
\Lhalpha \approx f n_e^2\,j_{H\alpha}\,V
\end{equation}

\noindent
with {$j_{H\alpha}=3.534 \times 10^{-25}$\ergcmcmcms} at $T=10000$K
\citep{Osterbrock_2006}. Therefore, the gas mass can be written as:

\begin{equation} \label{eq:Mg2}
M_g = \frac{m_p \Lhalpha }{n_e j_{H\alpha}}
\end{equation}

\begin{figure}
    \centering
	\includegraphics[width=\columnwidth]{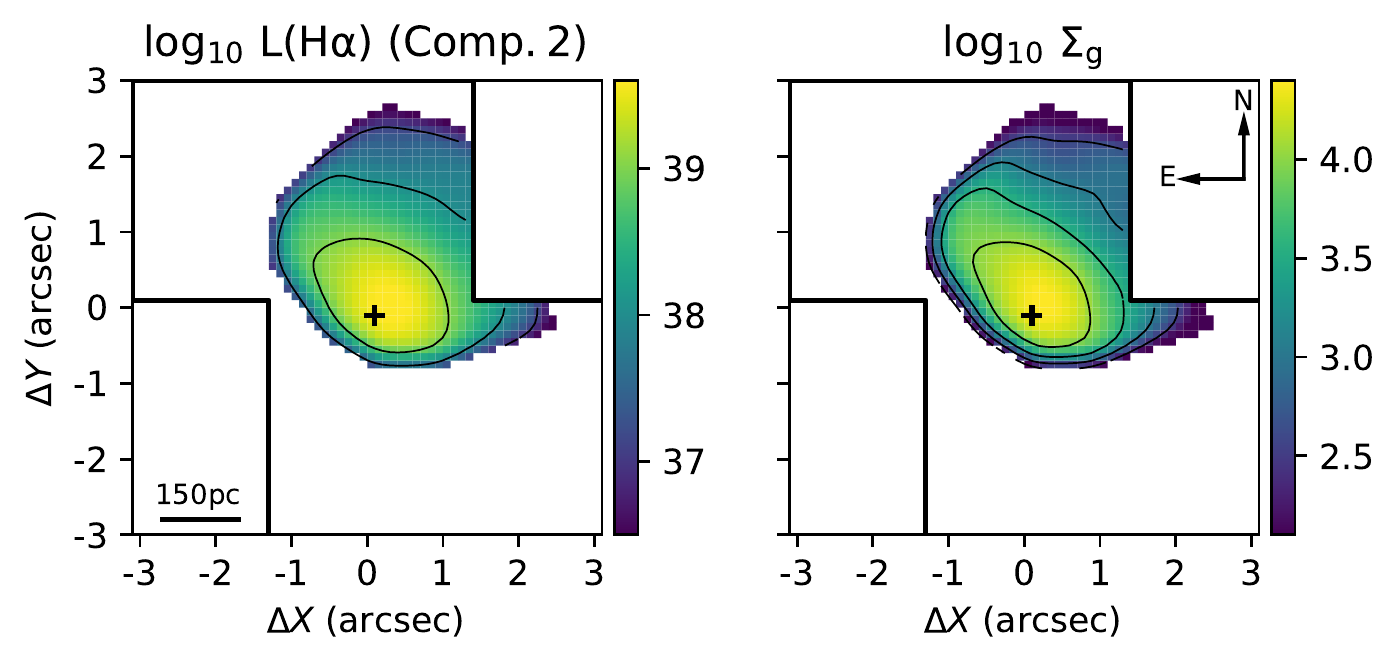}
    \caption{Left: Logarithm of Component 2 dereddened \halpha luminosity in erg s$^{-1}$. Right: Logarithm of Component 2 surface mass density in units of \msun ($0\farcs1$)$^{-2}$. Black cross is the peak of the continuum emission.}
    \label{fig:Outflow}
\end{figure}

\noindent
Our estimates for $n_e$ are based on the flux ratio between \sii lines, as described in \refappendix{sec:Av_ne}.
Please note the \Lhalpha used here is based on the flux of Comp. 2 \halpha emission only, corrected by extinction as described in \refappendix{sec:Av_ne}. 
The spatially resolved \Lhalpha and M$_{g}$ (i.e. gas mass surface density, $\Sigma_{\rm g}$) are shown in \autoref{fig:Outflow}.

The integrated gas mass is $M_g = 3.6 \pm 0.6 \times 10^{6}\msun$.
The t$_d$ was estimated as the ratio between the maximum extension of the outflow $\sim3\,\,{\rm arcsec} \approx 450\,\,{\rm pc}$ (see \autoref{fig:PV}) to its mean velocity, $\left \langle  v_{out} \right \rangle \approx 210$\kms, which gives t$_{\rm d} \approx 2.2 \times 10^{6}\,\,{\rm yr}$.
Combining these values we estimated an outflow mass rate of  \Mout $\approx$ 1.6 $\pm$ 0.6 \msun. 

We also estimated the outflow kinetic power as:

\begin{equation} \label{eq:OutPower}
\Eout = \frac{\Mout v_{out}^2}{2}
\end{equation}

\noindent
where $v_{out}$  is the radial velocity of Component 2 (\autoref{fig:Comp2}).
Integrating the entire outflow we obtained a total outflow kinetic power $\Eout \approx 2.2 \pm 0.3 \times 10^{40}\,\,{\rm erg\,s^{-1}}$.
In \autoref{sec:Agn_Wind}, we compare these values with the ones found in the literature and those expected by outflow-AGN relations.

 \section{Ionisation Mechanisms}
\label{sec:Ion_Mec}

\begin{figure*}
    \centering
	\includegraphics[width=0.75\textwidth]{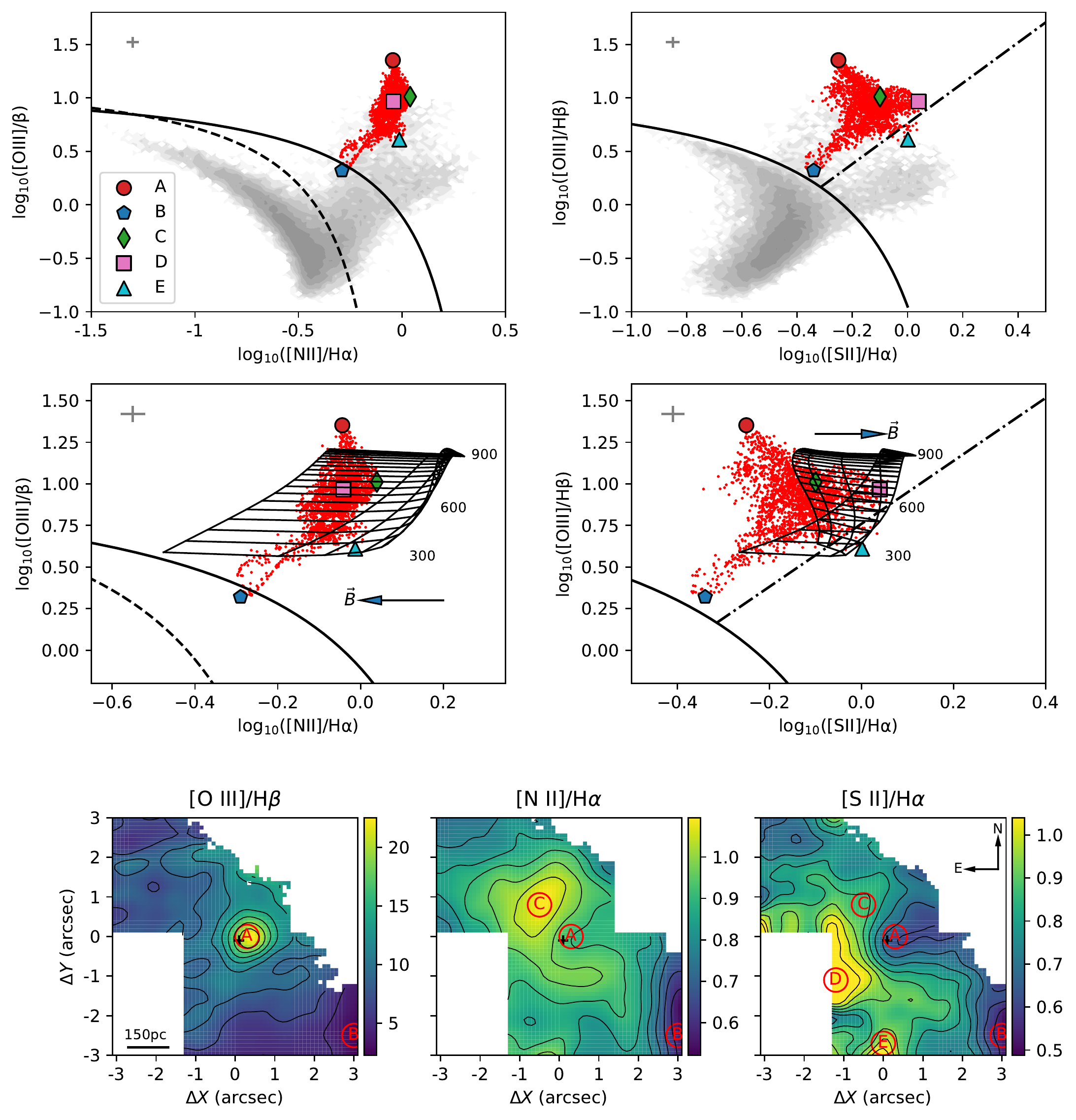}
    \caption{Top: \nii/\halpha vs. \oiii/\hbeta and
    \sii/\halpha vs. \oiii/\hbeta BPT diagrams; grey countours are
    galaxies from the SDSS main galaxy sample. red points are the spaxel of NGC~2992.
    Colored markers are the selected regions of NGC~2992, as shown in the bottom
    panels, representing: red (Region A), blue (B), green (C), pink (D) and cyan (E). Middle: Zoomed BPT diagrams with 
    shock+precursor grid models as measured with MAPPINGS V code \citep{Sutherland_2018}; the $\vec{B}$ indicates
    increasing magnetic field of the model (from 0.01 to 4
    $\mu$G); the black numbers indicates the velocities of the
    shocks, ranging from 300 to 900\kms. Bottom:
    Respectively, from left to right, \oiii/\hbeta,
    \nii/\halpha, and \sii/\halpha emission lines ratio maps;
    red circles show the selected regions shown in the upper
    panels. In BPT diagrams the grey crosses are the typical errors in the emission line ratios; the solid black lines are
    \citet{Kewley_01} theoretical upper limits for pure
    star forming galaxies. In the \nii/\halpha vs. \oiii/\hbeta
    diagram the dashed black line represents
    \citet{Kauffmann_03} line which traces the upper boundary
    of the SDSS star formation sequence. In the \sii/\halpha
    vs. \oiii/\hbeta diagram the dashed-dotted black line
    represents the empirical division between AGN like and
    low-ionisation nuclear emission-line region (LINER)
    \citep{Kewley_06}.}
    \label{fig:BPT}
\end{figure*}

Given the presence of young stars discussed in \autoref{sec:stellar}, X-ray (\autoref{sec:intro}) and BLR emission (\autoref{sec:kinematics}) which indicate AGN excitation, and the presence of radio emission that can drive shocks \citep{Dopita_02,allen_08}, we used the \nii/\halpha vs. \oiii/\hbeta and \sii/\halpha vs. \oiii/\hbeta \citet[][BPT]{Baldwin_81} diagnostic diagrams to investigate which of these excitation mechanisms are dominant in the innermost region of the galaxy.

In the bottom panels of  \autoref{fig:BPT}, we show the maps of \oiii/\hbeta, \nii/\halpha and \sii/\halpha line ratios for the entire line profile (by summing the kinematical components; we do that to increase the precision in the flux measurement and to avoid regions where a second component is detected in high S/N lines, like \oiii, but not detected in lower S/N ones, like \hbeta.)
We excluded the regions where uncertainties in the total flux, as shown in \refappendix{sec:Error}, are higher than 30 per cent.
The \oiii/\hbeta shows a clear AGN signature, with its peak, up to \oiii/\hbeta $\approx$ 25, at the centre of the FoV and decreasing radially down to  \oiii/\hbeta $\approx 2$ at SW corner of the FoV.
The \nii/\halpha and \sii/\halpha, however, do not show the same pattern: their maximum values are off-nuclear, and they do not present a radial decrease as in the \oiii/\hbeta one.
This indicates a lower energy mechanism (lower than an AGN, but higher than H~{\sc ii} regions) can be ionising the gas too.
In fact, shock ionised regions are characterised by having smaller \oiii/\hbeta ratios and larger \sii(\nii)/\halpha than those ionised by AGN \citep{Veilleux_87,Kewley_06, allen_08}. 

We selected limiting regions in the line ratio maps and shown it in the respectively BPT diagrams in the top panels of \autoref{fig:BPT}.
We also show some division lines: the solid black curves on both diagnostic diagrams trace the \citet{Kewley_01} theoretical upper limit of regions ionised by pure star formation (K01 line).
All spectra lying above the K01 line are dominated by ionisation mechanisms more energetic than star formation.
Several excitation mechanisms can increase the collisional excitation rate, enhancing the ratios of the forbidden to recombination lines, among them energetic photons from a power law-like emission (AGN); shock excitation and evolved stellar populations (pos-AGB) can also produce line ratios along this branch of the diagram.
The latter case does not apply to the nuclear region of NGC~2992 given that it requires EW(\halpha) < 3 $\textup{\AA}$\xspace  \citep{Stasi_ska_2008,Cid_Fernandes_2011} and our entire FoV has EW(\halpha) > 10 $\textup{\AA}$.
We also show the dashed black curve on the \nii/\halpha vs. \oiii/\hbeta diagnostic diagram: the \citet{Kauffmann_03} empirical classification line (Ka03 line), which traces the
upper boundary of the Sloan Digital Sky Survey \citet[SDSS][]{Strauss_02} star formation sequence.
All spectra lying below the Ka03 line are dominated by star formation.
The dashed-dotted black curve on the \sii/\halpha vs. \oiii/\hbeta  diagnostic diagram traces the \citet{Kewley_06} empirical classification line (K06 line), which separates high ionisation spectra associated with Seyfert (AGN) and other intermediary ionisation mechanisms including shock excitation.
We also show as grey points in the top BPT diagrams a collection of single-fibre emission line measurement from SDSS main galaxy sample by \citet{Kewley_06}.

Region A is the peak of the \oiii/\hbeta map, it shows clear AGN-like ratios, with an upper \oiii/\hbeta value higher than any SDSS AGN, which corresponds to a 3\arcsec diameter aperture, and therefore has a diluted AGN/host-galaxy emission line ratio. 
Region B has the lowest value in all the maps; it lies below K01, showing the radial decrease in the AGN contribution to the emission lines and significant contributions from both star formation and AGN excitation.
Regions C, D, and E display the highest values in the \sii/\halpha and \nii/\halpha maps; however, they have lower \oiii/\hbeta values than region A.
Region E, in fact, crosses the K06 line in the \sii/\halpha diagram, which again indicates the presence of another ionisation mechanism. 

In order to search for an explanation to the increase in this intermediary ionisation energy lines (\nii and \sii) we used the  \textsc{MAPPINGS V} shock models by \citet{Sutherland_2018}.
We explored the parameter space in the pure shock and shock+precursor grids of the models.
None of the pure shock models could explain these line ratios, given the high \oiii/\hbeta values.
However shock+precursor (Shock ionisation in a medium already ionised by power-law like emission) with shock velocities between 300\kms and 900\kms and magnetic field ($\vec{\rm B}$) between 0.01$\mu$G and 4$\mu$G, with a two-solar metallicity and a pre-shock density of 1 cm$^{-3}$,  successfully reproduce the line ratios of regions C, D and E.
In the middle panels of \autoref{fig:BPT} we show the grid models in these shock velocity and magnetic field ranges superimposed with the emission line ratios of the regions.
The shocks in NGC~2992 have been extensively studied by \citet{allen1999}, who conclude that shocks are the predominant ionisation mechanism in the extended NLR cones at a distance of several kpc from the nucleus.

We can conclude that a more simple Starburst-AGN mixture \citep[e.g.][]{Davies_14,Davies_14b,D_Agostino_2018} in which the maximum values of all \oiii/\hbeta, \nii/\halpha and \sii/\halpha are located in the innermost spaxel (location of the AGN), and all decrease radially towards the H~{\sc ii} ionisation region of the BPT in a straight line \citep[see for example figure 1 in][]{Davies_14b}, cannot describe the ionisation at the circumnuclear region of NGC~2992. Instead, a more complex Starburst-AGN-Shocks mixture is responsible for its ionisation.
However, the very high values of \oiii/\hbeta indicate the most important among these three mechanisms is AGN ionisation, at least in the innermost portion (see further discussion in \autoref{sec:multi_ion}).

\section{DISCUSSION}
\label{sec:discussion}

\subsection{Interaction-Driven Circumnuclear Star formation}
\label{sec:merger}


In general terms, the stellar population synthesis discussed in \autoref{sec:stellar} shows that stellar content in the inner 1.1~kpc of NGC~2992 is mainly composed by an old metal-rich population, with a smaller but considerable contribution from young metal-poor stars.
The presence of the latter is supported by other studies, for instance, using the \citet{1986A&A...162...21B} base of star clusters for stellar population synthesis, \citet{1990MNRAS.245..749S} found a predominance of old stellar populations (10$^{10}$ yr) and a small contribution of at least 5 per cent by recent star formation in the inner $5\arcsec$.
Moreover, \citet{Friedrich_2010} find that 10-20 per cent of the IR flux is attributable to a starburst, based on the emission from polycyclic aromatic hydrocarbon (PAH) molecules.
Additionally, these authors estimate the starburst to have occurred between 40-50~Myr, based on the equivalent width of the Br$\gamma$ line.

Major merger processes are known to be responsible for gas inflow towards the central regions of galaxies enhancing circumnuclear star formation and possibly triggering nuclear activity.
Merger galaxies have an increase in the SFR, mainly at the nuclear regions \citep{Ellison_2013, Pan_2019}. Thus, these sources have younger nuclear populations than those found in isolated galaxies, in which most of their recent formed stars are concentrated at the spiral arms.
Moreover, major mergers are also able to modify the metallicity gradients of galaxies \citep{Barrera-Ballesteros_2015}.
Galaxy simulations show that the gas, originally located at external portions of the isolated galaxy (metal-poor regions), moves towards internal regions (metal-rich regions) of the companion galaxy during the encounter and is capable of cooling and forming stars..
Such processes are able to explain both the increase in the SFR and the modifications in the metallicity gradient \citep{Dalcanton_2007,Torrey_2012,Sillero_2017}.
In fact, the pericentre passage between NGC~2992 and its companion NGC~2993 is estimated to have occurred $\sim$ 100 Myr ago \citep{Duc_2000}.
Thus, a possible scenario is that metal-poor gas inflow has led to interaction-driven circumnuclear star formation which can explain the presence of such young metal-poor stellar population in the nucleus.
Such inflows could also be responsible for triggering NGC~2992's nuclear activity, a scenario which will be further explored using numerical simulations in Paper~III.

\subsection{Feeding vs Feedback}

We can compare the mass outflow rate, estimated in \autoref{sec:Out_Prop}, with the accretion rate necessary to power the AGN at the nucleus of NGC~2992, calculated as follows:

\begin{equation}
\dot{M}_{\rm BH} = \frac{\Lbol}{c^2\eta}
\end{equation}

\noindent
where $\eta$ is the mass-energy conversion efficiency, which for Seyfert galaxies is usually assumed to be $\eta=0.1$ \citep{2002apa..book.....F}, \Lbol is the bolometric luminosity of the AGN, and $c$ is the speed of light.
To measure \Lbol we applied the bolometric correction by \citet{Marconi_04} on the most recent X-ray luminosity measurement of NGC~2992 made by \citet{Marinucci_2018}.
The author used a 2015 {\it NuSTAR} observation and obtained an absorption corrected 2-10 Kev Luminosity of $7.6 \pm 0.1 \times 10^{42}\,\,{\rm erg s^{-1}}$ that translates to $\Lbol = 9.1 \pm 0.2 \times 10^{43}\,\,{\rm erg s^{-1}}$.
We use these values to derive an accretion rate of $\dot{M}_{\rm BH} \approx 0.02\,\,\msun yr^{-1}$.

The nuclear accretion rate is two orders of magnitude smaller than the mass outflow rate (\Mout $\approx$ 1.6 $\pm$ 0.6 \msun yr$^{-1}$, see \autoref{sec:Out_Prop}). This implies that most of the outflowing gas does not originate in the AGN, but in the surrounding ISM, and this result supports the scenario in which the plasma bubbles (radio loops) are expanding, pushing gas away from the nuclear region.
However, the ratio between the kinetic power of this blueshifted outflow loop and the bolometric luminosity is $\sim 0.002$, which means only 0.2 per cent of the AGN radiative power is in this outflow.
Many authors argue that such a low value is not enough to have the feedback effects expected by cosmological simulations \citep{2005Natur.435..629S, 2012RAA....12..917S, 2017ARA&A..55..343B},
stating that \Eout/\Lbol should be closer to 5 per cent or above
in order to effectively suppress growth in the most massive galaxies.
However, some models show that coupling efficiencies of \Eout/\Lbol $\approx$ 0.5  per cent \citep{Hopkins_2010}, and even 0.1 per cent \citep{Costa2018} could successfully prevent star formation.

\subsection{AGN-Wind Scaling Relations}
\label{sec:Agn_Wind}

Comparing our outflow measurements with those from the literature we noted ours results disagree with those presented in \cite{M_sanchez_2011}, who reported \Mout= 120 M$_{\odot}$ yr$^{-1}$ and \Eout= 2.5$\times$10$^{42}$ erg s$^{-1}$, both almost one hundred times greater than ours.
We attribute this difference to two factors:
    \textbf{i)} they assumed the filling factor $f$ to be constant equal to 0.001, while we have measured the filling factor from L(\halpha) and $n_e$ (see \autoref{eq:Lha}). This assumption makes their M$_g$ proportional to $n_e$, and not inversely proportional  (as in \autoref{eq:Mg2}); 
    \textbf{ii)} they assume an electronic density of $n_e$ = 5000 cm$^{-3}$, a value more than 5 times larger than the maximum value we measured in the nuclear region of the galaxy.
These two factors combined explain their overestimation of both \Mout and \Eout.
We also argue these differences cannot be due to a larger coverage of the outflow by a larger field of view, in the sense of seeing the farthest part of the radio emission major loop and its redshifted outflow structure, given the authors have used the OSIRIS/VLT instrument, which has an even smaller FoV than GMOS.

We can compare our values to the properties of outflows found in other precious papers across the literature. In \autoref{fig:Outflow}, we show the
\cite{2017A&A...601A.143F}
compilation of a set of outflow energetics from various studies over the years spanning
a large range of AGN luminosity (\Lbol $\geq 10^{44}$), which they use to establish AGN wind scaling relations, mass outflow rate and outflow power as a function of bolometric luminosity, together with several measurements of nearby galaxies from individual studies compiled by us from the literature or obtained by our collaborators: NGC~1068, NGC~3783, NGC~6814,
NGC~7469 \citep{M_sanchez_2011},
NGC~4151 \citep{Crenshaw_2015},
Mrk~573 \citep{Revalski_2018},
Mrk~34 \citep{Revalski_2018}, 
NGC~7582 \citep{2009MNRAS.393..783R},
Mrk~1066 \citep{2010MNRAS.404..166R},
Mrk~1157 \citep{Riffel_2011}, 
Mrk~79 \citep{2013MNRAS.430.2249R},
NGC~5929 \citep{Riffel_2013},
NGC~5728 \citep{Shimizu_2019}, 
and NGC~3081 \citep{2014MNRAS.437.1708S},
ESO~362-G18 \citep{2018A&A...614A..94H},
NGC~1386 \citep{Lena_2015},
ESO~153-G20 \citep{2019MNRAS.489.4111S},
3C~33 \citep{2017MNRAS.469.1573C},
including NGC~2992 measurement by \cite{M_sanchez_2011}. 
Although these estimates are made using different recipes, leading to differences between methods, they are generally similar to the one adopted here.
Therefore, this comparison with the literature may be useful at least as an order of magnitude approximation.

As we can see, the overestimated previous measurement of NGC~2992's outflow made this source to be an outlier in the \cite{2017A&A...601A.143F} relations, while our values make the galaxy more likely to be explained by these relations.
However, the other measurements in nearby Seyfert galaxies support the existence of a larger scatter of these relations, both in \Eout and \Mout, at lower luminosities.
To confirm whether this extension is valid or not, more detailed studies like this, mainly using IFU, are needed so that outflow properties can be properly and systematically measured, without assuming fixed values of $n_e$ and $f$. 

\begin{figure}
    \centering
    \includegraphics[width=\columnwidth]{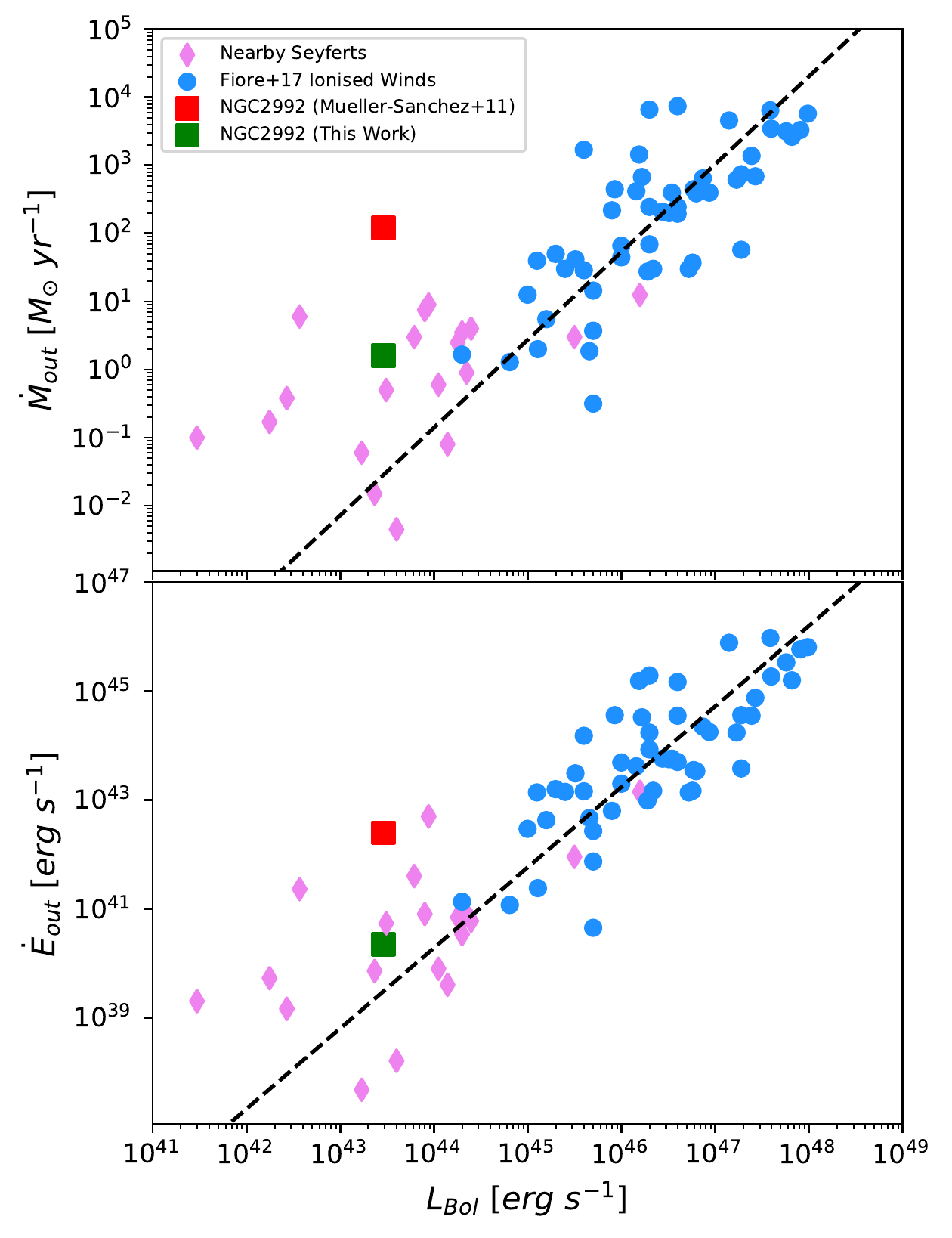}
    \caption{
        Correlations between the mass outflow rate (\Mout, Top Panel) and kinetic outflow power
        (\Eout, Bottom Panel).  Blue points are the data compiled by \citet{2017A&A...601A.143F} and the
        black dashed lines show the best fit correlations for these points. Red Squares are the previous measurement of
        NGC~2992 by \citet{M_sanchez_2011}, and green ones are the results of this work. Violet diamonds are
        compiled from various literature sources for more modest AGN luminosities.
    }
    \label{fig:OutflowBib}
\end{figure}

\subsection{Multiple Ionisation Mechanisms}  
\label{sec:multi_ion}

In \autoref{sec:Ion_Mec} we showed that despite being dominated by AGN excitation, star formation and shocks cannot be excluded as ionisation mechanisms in the nuclear region of the galaxy.
The existence of star formation ionisation in the nucleus is confirmed by \citet{Friedrich_2010} IR emission line analyses, as mentioned in \autoref{sec:merger}.
The presence of shocks is also supported by \citet{allen1999} using long-slit spectroscopy. The authors also argue the dominance of shocks increases, compared to ionisation by the central source, at larger radius.

Galaxies undergoing rapid phases of evolution (through processes such as galaxy-galaxy interactions, gas outflows and radio jets) often have multiple ionisation mechanisms contributing to their optical line emission \citep[e.g.][]{Rich_11,Lanz_2015,Rich_2015, Davies_2017}.
These multiple ionisation mechanisms are also in agreement with evolutionary scenarios in which gas inflows (driven either by secular mechanisms, like nuclear bars or nuclear spiral arms, or environmental ones, like minor and major mergers) triggers nuclear star formation and nuclear activity (feeding process) and then the active nucleus can drive both ionising photons from the accretion disk and radio jets (feedback process) that impacts the ISM physical conditions \citep{Hopkins_08,Hopkins_10}.

\section{SUMMARY AND CONCLUSIONS}
\label{sec:conclusion}

We have analysed the stellar population, the ionisation mechanism and kinematics of the ionised gas in the inner
$1.1\,\,{\rm kpc}$ of the interacting Seyfert galaxy NGC~2992, using optical spectra obtained with the GMOS integral
field spectrograph on the Gemini South telescope, with a spatial resolution of $\approx 120\,\,{\rm pc}$ and spectral
resolution $\approx 40\,\,{\rm km\,s^{-1}}$. The main results are:

\begin{itemize}
    \item
        The stellar population in the nuclear region of the galaxy is mainly  composed (60 per cent$ \leq x_O \leq 80$ per cent) by an old (t > 1.4 Gyr) metal-rich ($\meanZ_O \geq 2.0$\zsun) population with a smaller, but considerable contribution (10 per cent $\leq x_Y \leq 30$ per cent) of an young (t < 100 Myr) metal-poor ($\meanZ_Y \leq 1.0$ \zsun) population. A possible scenario is that metal-poor gas inflow during the pericentre passage of NGC~2993 has led to interaction-driven circumnuclear star formation, which can explain the presence of such young metal-poor stellar population in the nuclear region;

    \item
        The emission line analyses show the presence of both BLR and NLR emission, confirming its classification as an intermediate-type Seyfert galaxy. The BLR emission profile has a velocity dispersion of FWHM$_{\rm BLR} = 2010$\kms and a Balmer decrement of (\halpha/\hbeta)$_{\rm BLR} = 24.2$. The NLR presents two distinctly kinematical components: one that although disturbed is well fitted by a disk model, and therefore can be identified as gas in orbit in the galaxy disk, with radial velocities ranging from -130 to 130\kms; another that we interpreted as an outflow associated with the radio emission, with blueshifted velocities of $\sim$ 200\kms, a mass outflow rate of \Mout $\approx$ 1.6 $\pm$ 0.6 \msun and a kinematic power of \Eout $\approx$ 2.2 $\pm$ 0.3 $\times$ 10$^{40}$ erg s$^{-1}$;
    \item 
        The BPT diagnostic diagrams show the galaxy posses multiple ionisation mechanisms: a mixture of AGN, star-formation and shocks, the former being the dominant one. The \oiii/\hbeta ratio peaks at the innermost spaxels, and is located in the upper region of the BPT diagram. The lower values, located at the border of the FoV, decreases down to below the \citet{Kewley_01} line, showing an increase in the star-formation ionisation compared to the AGN ionisation at larger radius. Off-nuclear peaks in the \nii/\halpha and \sii/\halpha maps indicate the presence of another mechanism and are successfully explained by \citet{Sutherland_2018} shock+precursor models.

\end{itemize}

\section*{ACKNOWLEDGMENTS}
We thank the anonymous
referee whose comments helped us improve the methodology 
and the clarity of this work. G-P, M. thanks the Brazilian National Council for Scientific and Technological Development (CNPq) for his Master Scholarship and Taro Shimizu for providing a collection of data points for \autoref{fig:OutflowBib}. G.\,C. acknowledges the support by
the Comit\'{e} Mixto ESO-Chile and the DGI at University
of Antofagasta. 
J.\,A.\,H.\,J. thanks to Chilean institution CONICYT, Programa de Astronom\'ia, Fondo ALMA-CONICYT 2017, C\'odigo de proyecto 31170038. N.Z.D. acknowledges partial support from FONDECYT through project 3190769.
This work is based
on observations obtained at the Gemini Observatory, which
is operated by the Association of Universities for Research
in Astronomy, Inc., under a cooperative agreement with the
NSF on behalf of the Gemini partnership: the National Science
Foundation (United States), the Science and Technology
Facilities Council (United Kingdom), the National Research
Council (Canada), CONICYT (Chile), the Australian
Research Council (Australia), Minist\'erio da Ci\^encia e T\'ecnologia
(Brazil) and south-eastCYT (Argentina).

\section*{DATA AVAILABILITY}
The data underlying this article will be shared on reasonable request to the corresponding author.

\bibliographystyle{mnras}
\interlinepenalty=10000
\bibliography{references}

\appendix

\section{Stellar Populations Synthesis Maps}
\label{app1}

We show the maps of the recovered properties from three spatially resolved \textsc{starlight} synthesis we have performed, namely \textit{BC03}, \textit{BC03+FC} and \textit{M11+FC}. The top panels show, from left to right, the per cent contribution to the total light for young, intermediate and old SSPs ($x_Y$, $x_I$ and $x_O$). In the middle panels the light-weighted mean stellar metallicity for the three age groups, from left to right, $\meanZ_Y, \meanZ_I$ and $\meanZ_O$. In the bottom panel the \textit{V}-Band extinction ($A_V$), the per cent contribution to the total light from the FC component ($x_{agn}$) and the light-weighted mean stellar age, \meanAge, are shown, respectively, from left to right.  

 \begin{figure}
    \centering
	\includegraphics[width=\columnwidth]{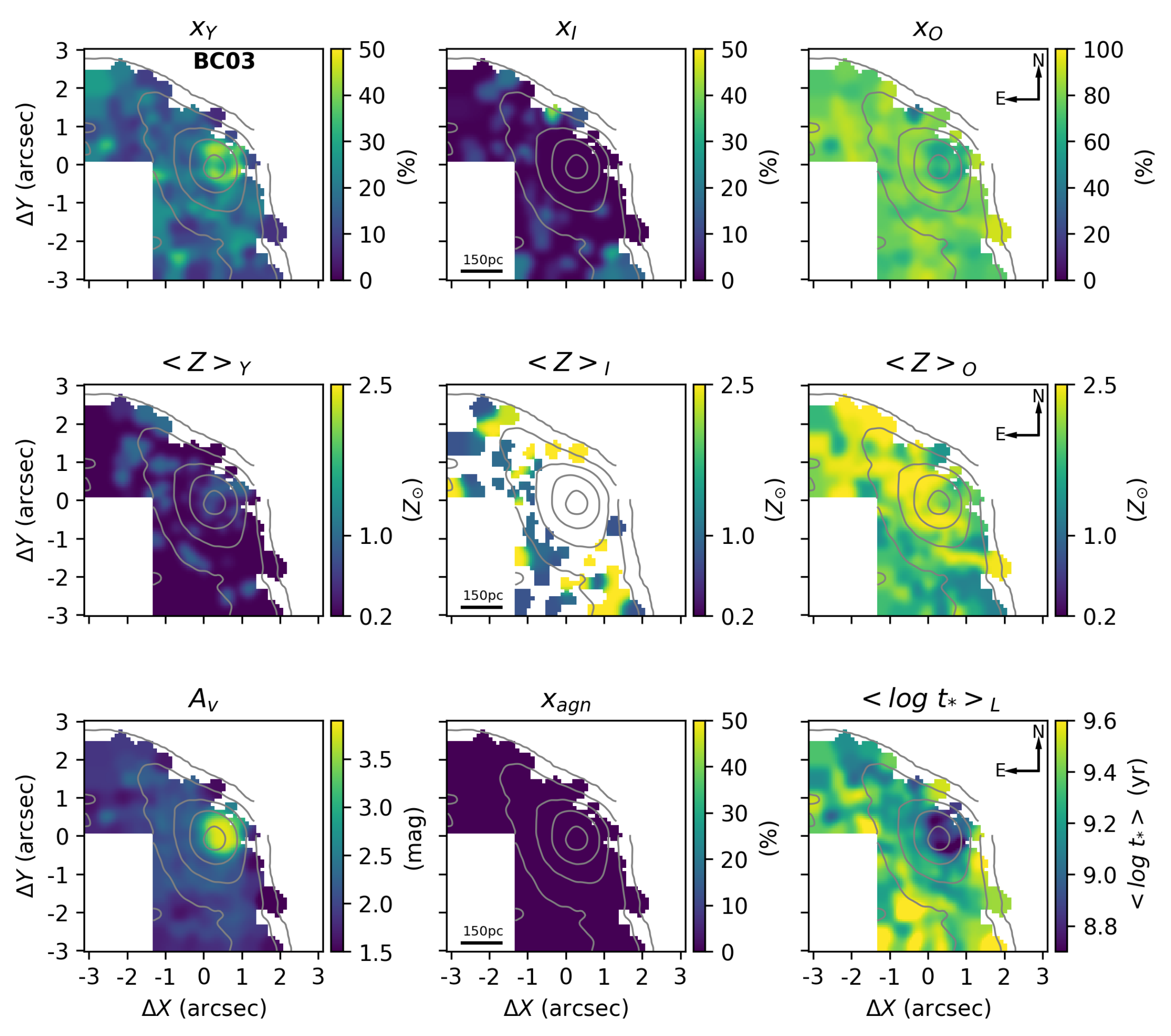}
     \caption{Maps of the recovered properties from the \textit{BC03} synthesis. Top: from left to right, percentage contribution to the total light from young, intermediate and old SSPs ($x_Y$, $x_I$ and $x_O$). Middle: from left to right, light-weighted mean stellar metallicity for the three SSP age groups,  $\meanZ_Y, \meanZ_I$ and $\meanZ_O$. Bottom left: $A_V$ extinction parameter. Bottom middle: Featureless continuum (FC) component percentage contribution to the total light ($x_{agn}$). Bottom Right: Light-weighted mean stellar age, \meanAge. The grey contour in all the maps is the continuum emission, left panel of \autoref{fig:Data}. The $x_{agn}$ is null in all points by definition, because there is no FC component in \textit{BC03} synthesis spectra base.}
     \label{fig:BC03}
\end{figure}

 \begin{figure}
    \begin{center}
        \centering
	    \includegraphics[width=\columnwidth]{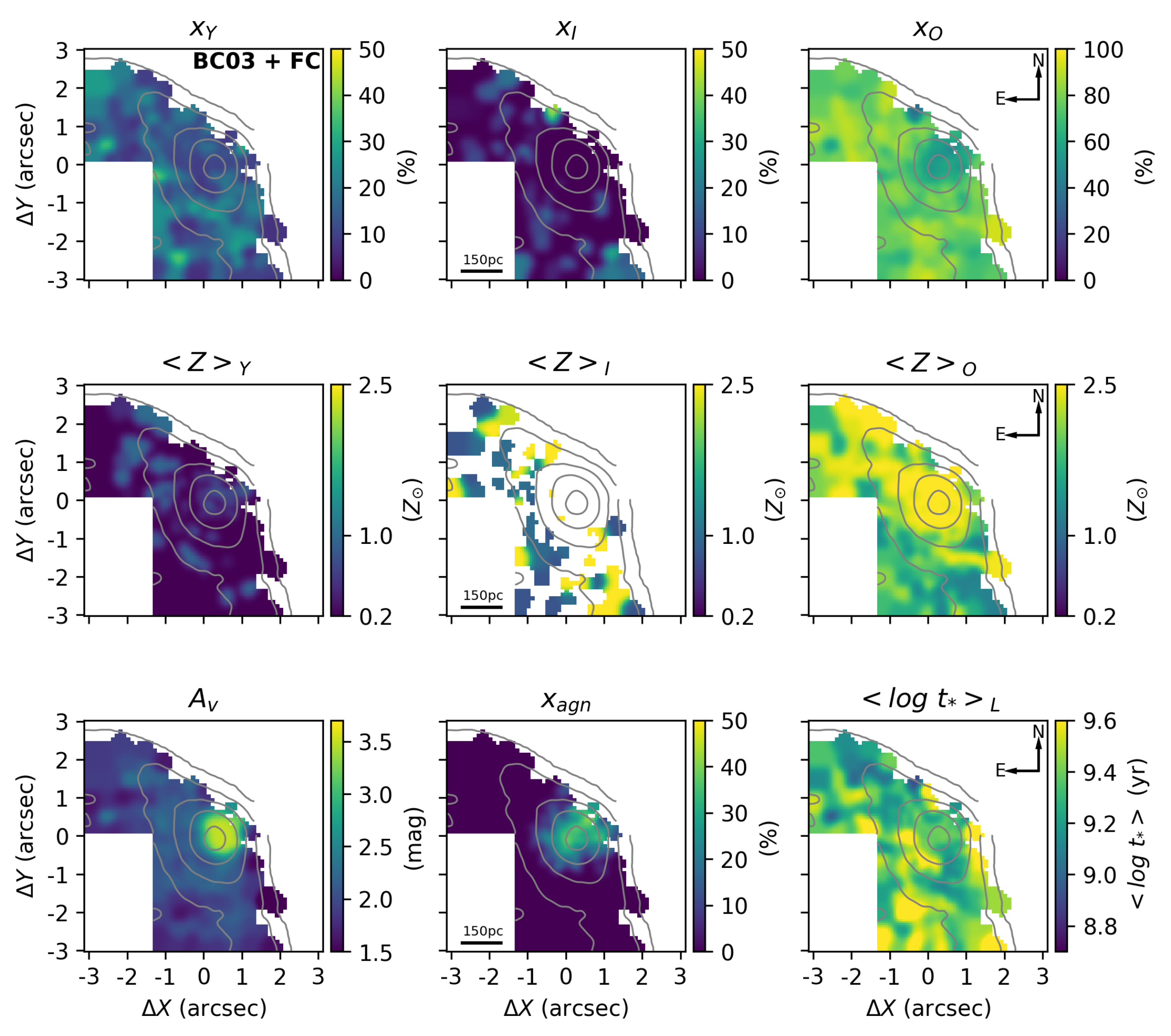}
        \caption{Same as \autoref{fig:BC03}for \textit{BC03+FC}.}
        \label{fig:BC03+FC} 
    \end{center}
\end{figure}

\begin{figure}
    \centering
	\includegraphics[width=\columnwidth]{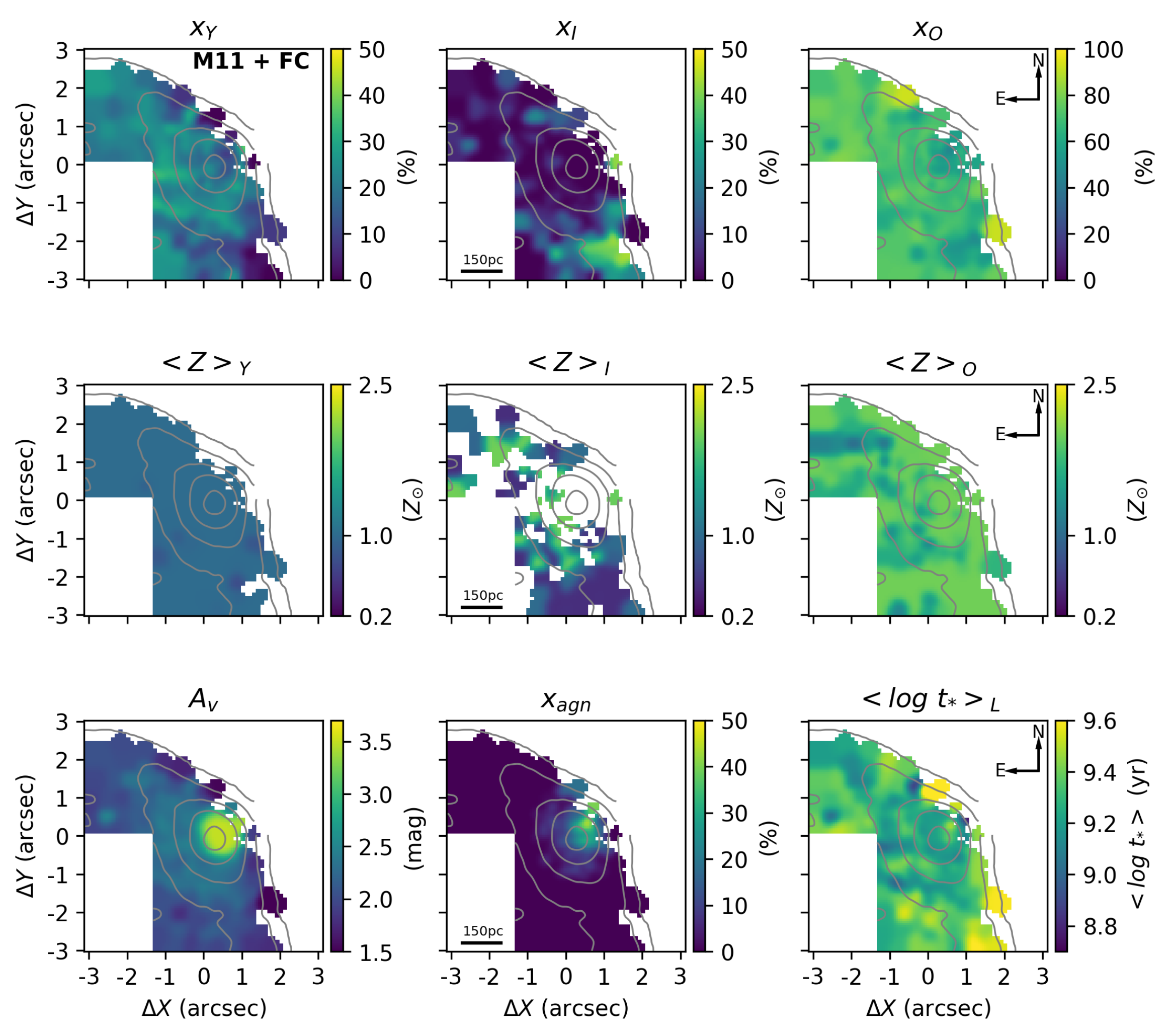}
     \caption{Same as \autoref{fig:BC03} for \textit{M11+FC}.}
     \label{fig:M11+FC}
\end{figure}

\section{Emission Line Fitting Uncertainties}
\label{sec:Error}

\begin{figure}
    \centering
	\includegraphics[width=\columnwidth]{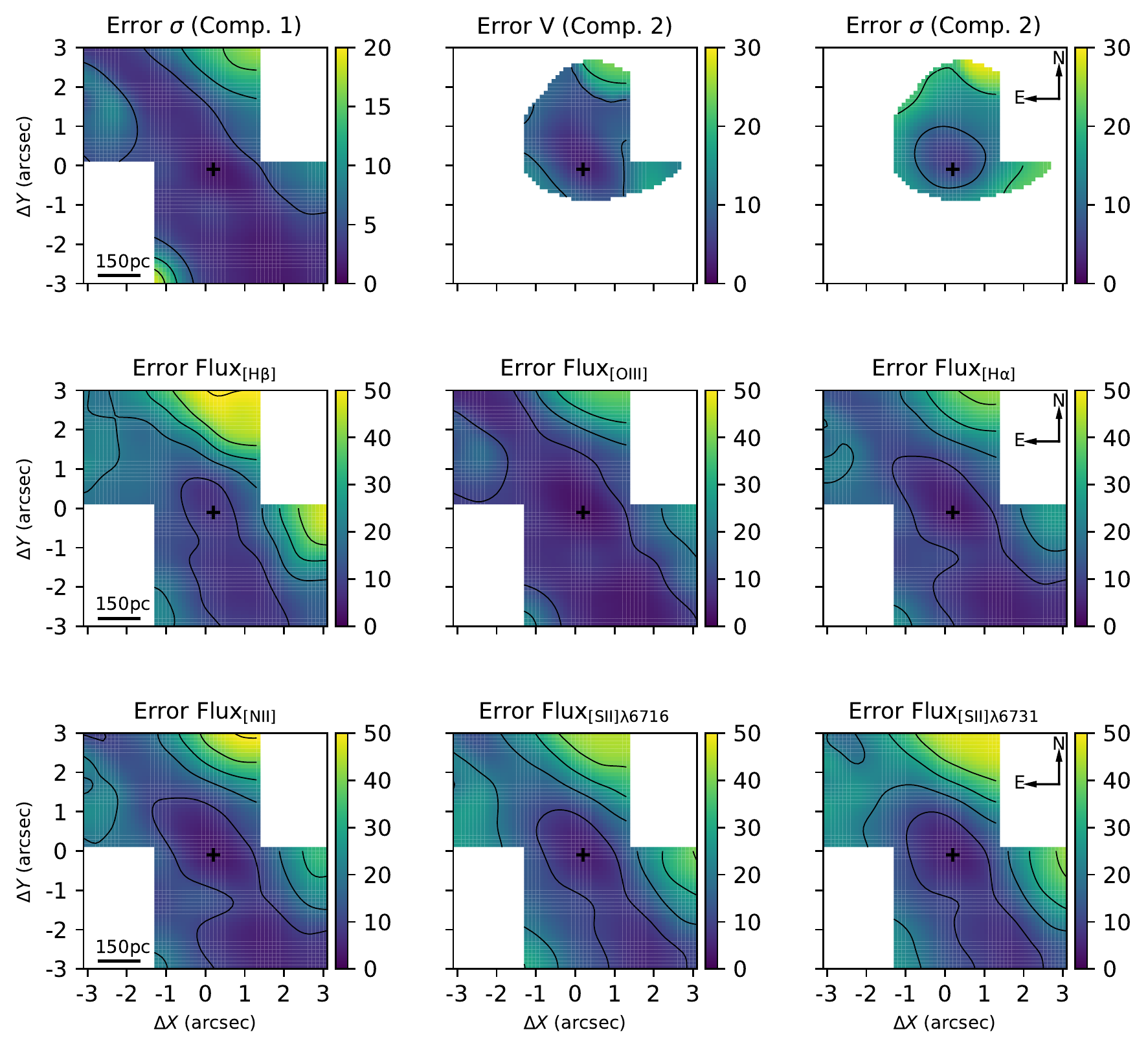}
     \caption{Emission line fitting uncertainties. Top: from right to left,
     velocity dispersion of Component 1 (\kms), radial velocity of Component 2
     (\kms), velocity dispersion of Component 2 (\kms). Middle and Bottom:
     percentage uncertainties in the integrated flux (Component 1 + Component) of \hbeta,
     \oiii$\lambda$5007, \halpha, \nii$\lambda$6583, and \sii$\lambda$6716,6731
     (\%). The black cross shows the peak of the continuum emission. Each contour in the maps means a 10\% variation in the uncertainties.}
     \label{fig:fit_err}
\end{figure}

We present here the uncertainties in the parameters of the emission lines fitting described in \autoref{sec:fit_EM}. The uncertainties are the standard deviation over the one hundred Monte Carlo iterations. The uncertainties in the radial velocity (V) of Component 1 was already presented in \autoref{fig:Comp1}. In \autoref{fig:fit_err} we show the uncertainties in the velocity dispersion ($\sigma$) of Component 1, the values are mostly smaller than 5\kms, reaching a maximum value of 15\kms.

In the top panel we also show the uncertainties of $v$ and $\sigma$ of Component 2, values are less than 10\kms in most the region, reaching up to 30\kms in a very small region, the $\sigma$ values are less then 10\kms in the peak of component emission and up to 30\kms in the borders of the detection region. In the middle and bottom panels the percentile uncertainties in the integrated flux (Component 1 + Component 2) of \hbeta, \oiii$\lambda$5007, \halpha, \nii$\lambda$6583, and \sii$\lambda$6716,6731 are shown.



The \hbeta line uncertainties range from 10 per cent at the centre of the FoV, up to 45 per cent in the dust lane region. \oiii and \halpha values range from less than 5 per cent up to 25 per cent at the dust lane region. \sii and \nii values range from less than 10 per cent up to 35 per cent at the dust lane region. 

\section{Nebular Extinction and Electron Density}
\label{sec:Av_ne}

In order to measure the extinction in the NLR, we adopt the \citet{Calzetti_2000} extinction law, and assume the case B recombination and the intrinsic \halpha/\hbeta value for the NLR \citep{Osterbrock_2006}, which combined leads to the following expression to the \textit{V}-band extinction:  

\begin{equation}
A_{\rm v} = 7.95 \times \log_{10} \left( \frac{\rm H\alpha/H\beta}{3.1} \right)
\end{equation}

\begin{figure}
    \centering
    \includegraphics[width=0.9\columnwidth]{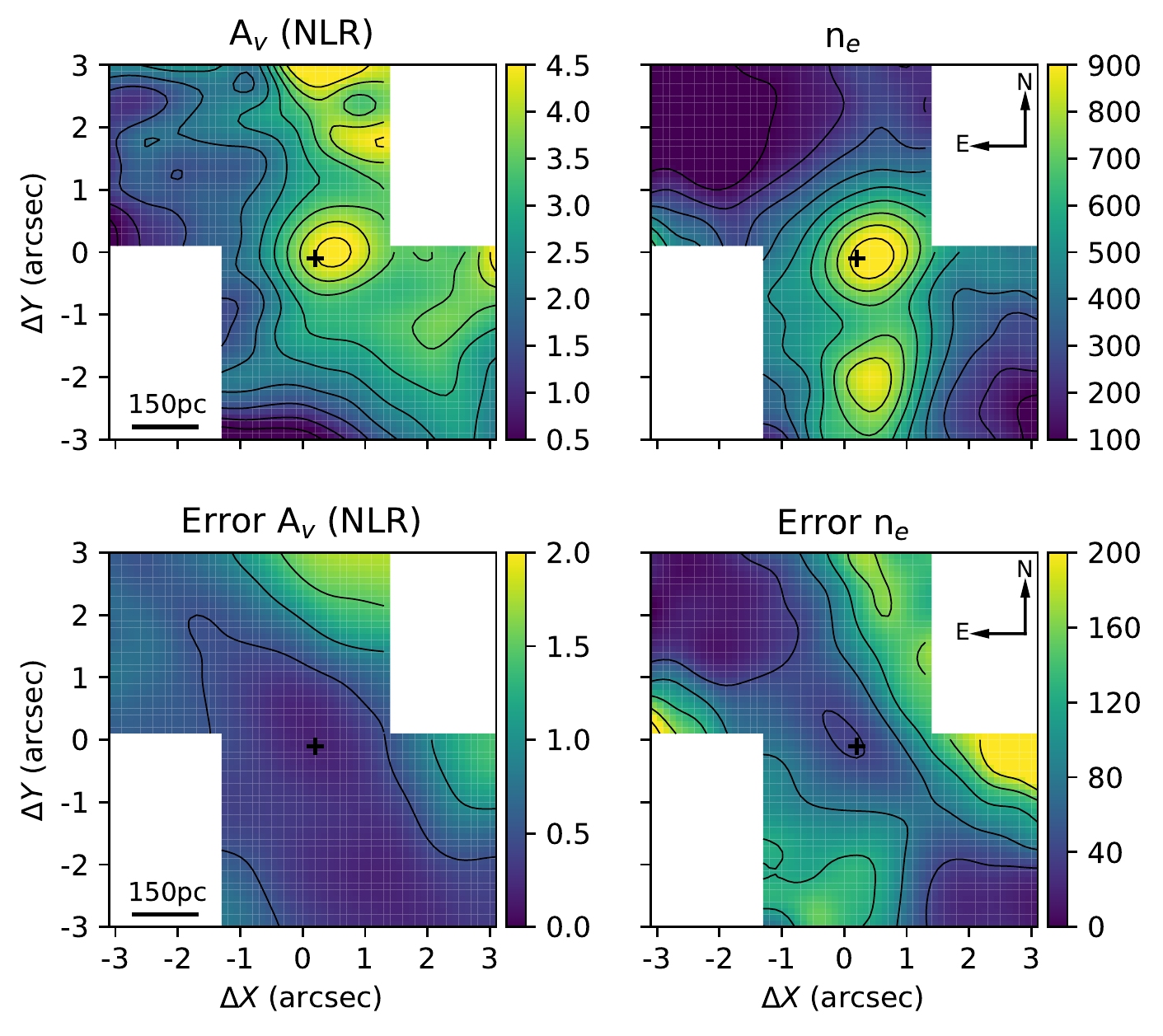}
    \caption{Top left and bottom left panels, shows, respectively the \textit{V}-band extinction $A_{\rm v}$ map and its uncertainties, both in units of magnitudes.  Top right and bottom right panel show, respectively, the electron density ($n_e$) and its uncertainties both in cm$^{-3}$ units. The black cross shows the peak of the continuum emission.}
    \label{fig:AV_ne}
\end{figure}

\noindent
In order to improve the S/N and given the fact that \hbeta Component 2 is detected in a very small region, we used the integrated flux (Component 1 + Component 2) of the NLR \halpha and \hbeta lines. The  $A_{\rm v}$ map is presented in the top left panel of \autoref{fig:AV_ne}, and its uncertainty in the bottom left panel. The $A_{\rm v}$ peak at $\sim$ 4.5 mag at the peak of the emission lines emission, and has values $>$ 3.5 mag in the dust lane portion of the FoV. However, this value may be even higher as the uncertainties are up to 2 mag in this region, caused by high impression on the \hbeta flux measurement, as seen in \autoref{fig:fit_err}. Although the $A_{\rm v}$ are not the same in the continuum \refappendix{app1} and in the NLR, the spatial patterns are similar: the higher values are found the central spaxels and towards the dust lane, and the lower towards the SE direction.

The electron density ($n_e$) of the ionised gas in the NLR can be obtained from the ratio between the \sii lines, $\sii\lambda$6716/$\lambda$6731 \citep{Osterbrock_2006}. We applied the solution of the collisional equilibrium equations given by \citet{Proxauf_2014} to obtain the $n_e$  values, as follows:

\begin{align}
\log_{10}(n_e \  [cm^3]) &= 0.0543\tan(-3.0553R + 2.8506) \nonumber \\
& + 6.98 - 10.6905R + 9.9186R^2 - 3.5442R^3
\end{align}

\noindent 
with R = $F_{6716}/F_{6731}$, the ratio between the \sii lines.
The $n_e$ map in its uncertainties are presented, respectively, in the top right and bottom right panel of \autoref{fig:AV_ne}. The maximum value of $n_e$ is 950 $cm^{-3}$ reached at the peak of the emission lines, decreasing radially down to 100 $cm^{-3}$, another high  ($\sim$ 800 $cm^{-3}$) density knot is present at the south of the nucleus. Both the values as well as the spatial profiles of both $A_{\rm v}$ and $n_e$ measured here are in agreement with the recent published analyses by \citet{Mingozzi_2019} using MUSE instrument, see their Figures A1 and B1.

\bsp	
\label{lastpage}
\end{document}